\documentclass{aa}  
\usepackage{graphicx}
\usepackage{txfonts}
\usepackage{xcolor}

\newcommand{\ros}{ROSAT}
\newcommand{\chan}{Chandra}
\newcommand{\xmm}{XMM-Newton}

\newcommand{\nh}{N_{\rm H}}
\def \msev{M7}

\def \magzs{\object{RX~J0720.4-3125}}
\def \magos{\object{RX~J1605.3+3249}}
\def \magot{\object{RX~J1308.6+2127}}

\def \magze{\object{RX~J0806.4-4123}}

\def \jsix{J1605}

\def \hess1731{\object{HESS~J1731-347}}
\def \pir14{P14}
\begin{document} 
\title{A deep \xmm\ look on the thermally emitting\\isolated neutron star \magos
\thanks{Based on observations obtained with \xmm, an ESA science mission with instruments and contributions directly funded by ESA Member States and NASA (Target \magos, large programme 76446; archival data 0073140201, 0073140301, 0073140501, 0157360401, 0671620101).}}
\author{A.~M.~Pires\inst{1}
   \and A.~D.~Schwope\inst{1}
   \and F.~Haberl\inst{2}
   \and V.~E.~Zavlin\inst{3}
   \and C.~Motch\inst{4}
   \and S.~Zane\inst{5}}
\offprints{A. M. Pires}
\institute{Leibniz-Institut f\"ur Astrophysik Potsdam (AIP), An der Sternwarte 16, 14482 Potsdam, Germany, 
   \email{apires@aip.de} 
   \and
   Max-Planck-Institut f\"ur extraterrestrische Physik, Giessenbachstra\ss e, 85748 Garching, Germany
   \and
   Universities Space Research Association, Science \& Technology Institute, Huntsville, AL 35805, USA   
   \and
   CNRS, Universit\'e de Strasbourg, Observatoire Astronomique, 11 rue de l'Universit\'e, F-67000 Strasbourg, France
   \and 
   Mullard Space Science Laboratory, University College London, Holmbury St. Mary, Dorking, Surrey, RH5 6NT, UK
}
\date{Received ...; accepted ...}
\titlerunning{A deep \xmm\ look on \magos}
\authorrunning{A.~M.~Pires et al.}
\abstract
{Previous \xmm\ observations of the thermally emitting isolated neutron star \magos\ provided a candidate for a shallow periodic signal and evidence of a fast spin down, which suggested a high dipolar magnetic field and an evolution from a magnetar.
We obtained a large programme with \xmm to confirm its candidate timing solution, understand the energy-dependent amplitude of the modulation, and investigate  the spectral features of the source.
We performed extensive high-resolution and broadband periodicity searches in the new observations, using the combined photons of the three EPIC cameras and allowing for moderate changes of pulsed fraction and the optimal energy range for detection. We also investigated the EPIC and RGS spectra of the source with unprecedented statistics and detail.
A deep $4\sigma$ upper limit of $1.33(6)\%$ for modulations in the relevant frequency range conservatively rules out the candidate period previously reported. Blind searches revealed no other periodic signal above the $1.5\%$ level $(3\sigma$; $P>0.15$\,s; $0.3-1.35$\,keV) in any of the four new observations. While theoretical models fall short at physically describing the complex energy distribution of the source, best-fit X-ray spectral parameters are obtained for a fully or partially ionized neutron star hydrogen atmosphere model with $B=10^{13}$\,G, modified by a broad Gaussian absorption line at energy $\epsilon=385\pm10$\,eV. The deep limits from the timing analysis disfavour equally well-fit double temperature blackbody models where both the neutron star surface and small hotspots contribute to the X-ray flux of the source.
We identified a low significance ($1\sigma$) temporal trend on the parameters of the source in the analysis of RGS data dating back to 2002, which may be explained by unaccounted calibration issues and spectral model uncertainties. The new dataset also shows no evidence of the previously reported narrow absorption feature at $\epsilon\sim570$\,eV, whose possible transient nature disfavours an atmospheric origin.
}

\maketitle
\section{Introduction\label{sec_intro}}
In the usual scenario of magnetic dipole braking in vacuum, the observed secular lengthening of a pulsar spin period, typically of 3\,s every $10^8$\,yr, is the consequence of the torque exerted by the magnetic field on the rotating neutron star. Despite its simplicity, the model provides useful estimates of the evolutionary state of a pulsar, namely, its available rotational power, characteristic age, and surface dipolar magnetic field strength \citep{1969ApJ...157.1395O}. For the isolated neutron stars (INSs) in our Galaxy, dipolar field estimates span across five orders of magnitude \citep{2005AJ....129.1993M}. Among the sources with the highest values are those known as magnetars \citep[see][for recent reviews]{2015RPPh...78k6901T,2017ARA&A..55..261K,2018MNRAS.474..961C,2018arXiv180305716E}. 

Magnetars are usually observed through their violent bursts of high energy and slow down at a much faster rate than normal pulsars (up to 1\,s every 60 years, for the extreme case of the soft-gamma repeater \object{SGR~1806-20}). According to the most favoured interpretation \citep{1995MNRAS.275..255T,1996ApJ...473..322T}, their complex phenomenology, transient behaviour, and bright quiescent X-ray luminosity, much in excess of that from spin down, can be explained by crustal and magnetospheric effects provoked by the decay and rearranging of the enormous stellar magnetic field \citep{1992ApJ...395..250G,2009A&A...496..207P,2016ApJ...833..261B}. 
As a result of field dissipation and braking, it is expected that an evolved magnetar ($\gtrsim10^5$\,yr) will be less active than young ones, and have a longer spin period and higher surface temperature than ordinary pulsars of similar age \citep[e.g.][]{2011ApJ...727L..51P,2011ApJ...740..105T}.

The group of X-ray thermally emitting INSs discovered by \ros\ and dubbed the magnificent seven \citep[\msev, see][for reviews]{2007Ap&SS.308..181H,2008AIPC..983..331K,2009ASSL..357..141T} may have evolved from such a channel of pulsar evolution \citep[e.g.][]{1998ApJ...506L..61H,2009ApJ...705..798K,2010MNRAS.401.2675P,2013MNRAS.434..123V}. 
They consist of a rather unique local group of middle-aged ($\sim10^5-10^6$\,yr), cooling neutron stars, displaying similar low blackbody temperatures ($kT\sim45-100$\,eV), long spin periods ($P\sim3-17$\,s), and moderately strong dipolar magnetic field strengths ($B_{\rm dip}\sim{\rm few}\times10^{13}$\,G). Unlike other X-ray pulsars, their emission is purely thermal with no signs of magnetospheric activity and is believed to originate directly from the neutron star surface. 

\begin{table}[t]
\caption{Log of the \xmm\ AO14 and 2012 observations of \magos
\label{tab_dataexposure}}
\centering
\begin{tabular}{@{}c c c c r r@{}}
\hline\hline
\textsf{\small obsid} & Date & Inst. & Mode & \multicolumn{1}{c}{Duration} & GTI \\
 & & & & \multicolumn{1}{c}{(s)} & (\%) \\
\hline
0764460201 & 2015-07-21 & {\small pn} & {\small FF} & 118\,344 & 97 \\ 
 & 2015-07-21 & {\small MOS1} & {\small LW} & 110\,089 & 100 \\ 
 & 2015-07-21 & {\small MOS2} & {\small LW} & 119\,079 & 100 \\ 
 & 2015-07-21 & {\small RGS1} & {\small SES} & 120\,169 & 100 \\ 
 & 2015-07-21 & {\small RGS2} & {\small SES} & 120\,113 & 100 \\ 
0764460301 & 2015-07-26 & {\small pn} & {\small FF} & 65\,038 & 96 \\ 
 & 2015-07-26 & {\small MOS1} & {\small LW} & 66\,650 & 100 \\ 
 & 2015-07-26 & {\small MOS2} & {\small LW} & 66\,629 & 100 \\ 
 & 2015-07-26 & {\small RGS1} & {\small SES} & 66\,858 & 100 \\ 
 & 2015-07-26 & {\small RGS2} & {\small SES} & 66\,757 & 100 \\ 
0764460401 & 2015-08-20 & {\small pn} & {\small FF} & 68\,656 & 81 \\ 
 & 2015-08-20 & {\small MOS1} & {\small LW} & 70\,459 & 91 \\ 
 & 2015-08-20 & {\small MOS2} & {\small LW} & 71\,629 & 90 \\ 
 & 2015-08-20 & {\small RGS1} & {\small SES} & 71\,858 & 91 \\ 
 & 2015-08-20 & {\small RGS2} & {\small SES} & 71\,777 & 89 \\ 
0764460501 & 2016-02-10 & {\small pn} & {\small FF} & 59\,932 & 96 \\ 
 & 2016-02-10 & {\small MOS1} & {\small LW} & 61\,540 & 100 \\ 
 & 2016-02-10 & {\small MOS2} & {\small LW} & 61\,495 & 100 \\ 
 & 2016-02-10 & {\small RGS1} & {\small SES} & 61\,749 & 100 \\ 
 & 2016-02-10 & {\small RGS2} & {\small SES} & 61\,675 & 100 \\ 
\hline
0671620101 & 2012-03-06 & {\small pn} & {\small FF} & 58\,542 & 68 \\ 
 & 2012-03-06 & {\small MOS1} & {\small FF} & 57\,432 & 89 \\ 
 & 2012-03-06 & {\small MOS2} & {\small FF} & 57\,008 & 90 \\ 
 & 2012-03-06 & {\small RGS1} & {\small SES} & 60\,406 & 88 \\ 
 & 2012-03-06 & {\small RGS2} & {\small SES} & 60\,406 & 84 \\ 
\hline
\end{tabular}
\tablefoot{The EPIC cameras were operated in imaging mode and the thin filter was used. The RGS detectors were operated in \textit{high event rate with SES} spectroscopy mode for readout. We list the percentage of good-time intervals (GTIs) after filtering out periods of high background activity (see the text for details).
}
\end{table}
The source \object{\magos}, as the third brightest among the \msev\ INSs \citep{1999A&A...351..177M}, was consequently visited by \xmm\ in several occasions during the early years of its science operations \citep[see][for details on the past investigations of the source]{2004ApJ...608..432V,2007Ap&SS.308..181H}. However, these early observations were not deep enough to allow the detection of the neutron star spin signal. Ensuing a visibility gap of six years, a 60\,ks observation performed in 2012 finally provided a candidate spin period for the INS \citep[][]{2014A&A...563A..50P}. The amplitude of the shallow and strongly energy-dependent periodic signal was detected close to the sensitivity limit of the data; only the harder portion of the source spectrum (roughly, 30\% of all source events at energies above $0.5$\,keV) was found to show a significant modulation at the $4\sigma$ level. Nonetheless, a joint timing analysis around the detected signal at $P\sim3.39$\,s, connecting the 2012 dataset with early \xmm\ observations of the source, hinted at an unprecedentedly high value of spin down. The inferred dipolar magnetic field of $B_{\rm dip}\sim7.4\times10^{13}$\,G overlaps the magnetar range and, if confirmed, could rank the highest in the group.

Besides, the analysis of the then available \xmm\ EPIC data on the source confirmed the evidence of a complex, multi-temperature, energy distribution and the presence of absorption features. In \citet{2014A&A...563A..50P} we described the spectrum of the source using a two-component blackbody model of temperatures $60$\,eV and $110$\,eV, superposed by one or two Gaussian absorption features at around energies $400$\,eV and $860$\,eV; best fits were found assuming the Galactic column density in the direction of the source, $2.4\times10^{20}$\,cm$^{-2}$ \citep[][see Section~\ref{sec_spec}, for details]{2005A&A...440..775K}. These results motivated us to investigate the INS further.

We were granted a large programme of observation with \xmm\ (programme ID: 76446) for a total duration of 310\,ks and four satellite visits in the AO14 observing cycle. 
We report here the results of this observational campaign. The paper is organised as follows: in Section~\ref{sec_data} we describe the \xmm\ observations and the data reduction. Analysis and results are presented in Section~\ref{sec_analysis}. The implications of our results are discussed in Section~\ref{sec_discussion}, with particular emphasis on the properties and recent work on the group of \msev\ INSs. Conclusions and the summary of results are in Section~\ref{sec_summary}.

\begin{table}[t]
\caption{Summary of archival \xmm\ observations of \magos\ used in the RGS spectral analysis
\label{tab_rgsdata}}
\centering
\begin{tabular}{c c c c c}
\hline\hline
Ref. & \textsf{\small obsid} & Date & \multicolumn{2}{c}{Net exposure (ks)} \\
\cline{4-5}
 & & & RGS1 & RGS2\\
\hline
(A) & 0073140201 & 2002-01-15 & 27.5 & 27.0 \\ 
(B) & 0073140301 & 2002-01-09 & 18.6 & 17.0 \\ 
(C) & 0073140501 & 2002-01-19 & 22.0 & 21.5 \\ 
(D) & 0157360401 & 2003-01-17 & 29.3 & 28.5 \\ 
2012 & 0671620101 & 2012-03-06 & 53.4 & 50.8 \\ 
\hline
\multicolumn{3}{l}{Total archival RGS data (ks)} & 151 & 145 \\
\multicolumn{3}{l}{Total RGS data\tablefootmark{$\star$} (ks)} & 465 & 457 \\
\hline
\end{tabular}
\tablefoot{
\tablefoottext{$\star$}{Taking into account the AO14 campaign.}}
\end{table}

\section{Observations and data reduction\label{sec_data}}
\begin{table*}[t]
\caption{Parameters of \magos, as extracted from the EPIC images in the AO14 and 2012 observations
\label{tab_sourceparam}}
\centering
\begin{tabular}{l r r r r r}
\hline\hline
Parameter\,/\,\textsf{\small obsid}  & 0764460201   & 0764460301  & 0764460401 & 0764460501 & 0671620101\tablefootmark{$\star$} \\
\hline
Detection likelihood& $2.5\times10^6$     & $1.4\times10^6$       & $1.2\times10^6$      & $1.2\times10^6$      & $0.9\times10^6$ \\
Counts              & $4.360(7)\times10^5$& $2.458(5)\times10^5$  & $2.185(5)\times10^5$ & $2.262(5)\times10^5$ & $1.669(5)\times10^5$ \\
\ldots$0.2-0.5$\,keV& $2.579(5)\times10^5$& $1.456(4)\times10^5$  & $1.305(4)\times10^5$ & $1.331(4)\times10^5$ & $0.987(3)\times10^5$ \\
\ldots$0.5-1.0$\,keV& $1.669(4)\times10^5$& $0.939(3)\times10^5$  & $0.823(3)\times10^5$ & $0.872(3)\times10^5$ & $0.6356(27)\times10^5$ \\
\ldots$1.0-2.0$\,keV& $1.11(12)\times10^4$& $0.625(9)\times10^4$  & $0.566(8)\times10^4$ & $0.586(9)\times10^4$ & $0.462(8)\times10^4$ \\
Rate (s$^{-1}$)     & $4.278(7)$          & $4.326(9)$            & $4.226(10)$          & $4.351(10)$          & $4.379(12)$ \\
\multicolumn{6}{l}{Rate (frame$^{-1}$\,camera$^{-1}$)} \\
\ldots pn   & $0.2832(4)$  & $0.2861(6)$  & $0.2804(6)$  & $0.2861(6)$  & $0.2842(7)$ \\
\ldots MOS1 & $0.5601(21)$ & $0.5992(28)$ & $0.5283(27)$ & $0.5921(29)$ & $1.589(10)$\tablefootmark{$\dagger$} \\
\ldots MOS2 & $0.6338(22)$ & $0.6328(9)$  & $0.639(3)$   & $0.641(3)$   & $1.889(11)$\tablefootmark{$\dagger$} \\
\tablefootmark{$\ddag$}HR$_1$ & $-0.2142(15)$  & $-0.2160(21)$  & $-0.2262(23)$ & $-0.2082(22)$ & $-0.2164(26)$ \\
\tablefootmark{$\ddag$}HR$_2$ & $-0.8757(13)$  & $-0.8753(17)$  & $-0.8714(18)$ & $-0.8740(18)$ & $-0.8644(21)$ \\
\tablefootmark{$\ddag$}HR$_3$ & $-0.9975(13)$  & $-0.9958(21)$  & $-0.9993(16)$ & $-0.9994(12)$ & $-0.9955(26)$ \\
RA (h min sec) & $16$\ \ $05$\ \ $18.5(6)$  & $16$\ \ $05$\ \ $18.5(6)$  & $16$\ \ $05$\ \ $18.5(7)$  & $16$\ \ $05$\ \ $18.5(8)$  & $16$\ \ $05$\ \ $18.4(9)$ \\
DEC (d m s) & $+32$\ \ $49$\ \ $19.3(5)$ & $+32$\ \ $49$\ \ $19.2(5)$ & $+32$\ \ $49$\ \ $19.7(6)$ & $+32$\ \ $49$\ \ $19.6(7)$ & $+32$\ \ $49$\ \ $18.7(8)$ \\
RA offset ($''$)  & $-0.6\pm0.4$ & $+0.2\pm0.3$ & $-0.7\pm0.4$ & $-0.2\pm0.5$ & $-1.2\pm0.3$ \\
DEC offset ($''$) & $+1.1\pm0.3$ & $+1.0\pm0.3$ & $+0.5\pm0.4$ & $+0.9\pm0.5$ & $-0.3\pm0.5$ \\
Reference sources & $56$ & $53$ & $50$ & $49$ & $38$ \\
\hline
\end{tabular}
\tablefoot{Counts and rates are given in the total \xmm\ energy band ($0.2-12$\,keV), unless otherwise specified. The EPIC source coordinates RA and DEC are astrometrically corrected, using as reference the GSC~2.3.2 catalogue (see text). The corresponding $1\sigma$ errors take into account the astrometric errors in each coordinate.
\tablefoottext{$\star$}{The source parameters, as extracted from the 2012 observation (\textsf{obsid} 0671620101), are shown for comparison.}
\tablefoottext{$\dagger$}{The 2012 MOS observations were operated in full-frame mode, yielding higher counts per frame in comparison with the AO14 MOS exposures.}
\tablefoottext{$\ddag$}{Hardness ratios (HR) are ratios between the difference and total counts in two contiguous of the first four \xmm\ energy bands.}}
\end{table*}

The \xmm\ observatory \citep{2001A&A...365L...1J} targeted the INS \magos\ (hereafter \jsix) in four occasions between July 2015 and February 2016, using EPIC as the prior instrument for the investigation. Table~\ref{tab_dataexposure} contains information on the scientific exposures and instrumental configuration of the EPIC-pn \citep{2001A&A...365L..18S}, EPIC-MOS \citep{2001A&A...365L..27T}, and RGS \citep{2001A&A...365L...7D} detectors. 

We included in the timing and spectral analysis (Section~\ref{sec_analysis}) the past 2012 observation of the source (\textsf{\small obsid} 0671620101; Table~\ref{tab_dataexposure}). 
For the spectral analysis of RGS data (Section~\ref{sec_rgsspec}), we additionally included archival \xmm\ observations of the source dating back to 2002 that were not severely affected by background flares (see Table~\ref{tab_rgsdata}). We applied as criterion a minimum net exposure of 10\,ks to include observations in the RGS analysis. The archival EPIC observations of \jsix\ performed before the visibility gap (analysed and discussed in \citealt{2014A&A...563A..50P} and references therein) were not included due to the high background level and the heterogeneous science operating modes and optical blocking filters that were then adopted (see also Section~\ref{sec_rgsspec}). 
All archival observations were processed and analysed consistently with the data from the large programme. 

The main trigger behind our programme was to confirm the candidate spin period and, by means of a precise timing solution, measure the neutron star's spin down rate. This is better achieved through a well-sampled ephemeris; incoherent methods result into much less accurate spin down determinations, while a two-dimensional periodicity search (like the one performed in \citealt{2014A&A...563A..50P}) suffers from a large parameter space and number of independent trials that have to be covered in the case of observations that are years apart. As a result, the pulsar best ($P,\dot{P}$) solution is determined at a low confidence level. 

Therefore, the time intervals between the four observations in AO14 (of $5$, $25\pm3$, and $175\pm8$ days) were carefully chosen to coherently connect them in phase with three past pn observations of the source, for a total time span of $\sim5160$ days (Section~\ref{sec_timing}). Assuming the properties of the periodic signal as detected in the 2012 observation (\citealt{2014A&A...563A..50P} and Section~\ref{sec_timing}), we considered for the feasibility a total count rate of 0.8\,s$^{-1}$ in the three EPIC cameras for source photons with energy above $0.5$\,keV; the individual exposure times were required to ensure a significant detection of the 2012 signal while allowing for moderate changes of pulsed fraction, $p_{\rm f}=2.5\%-5\%$ (within $\pm1\sigma$). Likewise, we chose to operate the MOS and pn cameras in large-window (LW) and full-frame (FF) imaging modes, to provide sufficient time resolution (0.9\,s and 73.4\,ms, respectively) to measure the 2012 modulation. We adopted the thin filter for both instruments, given its better response at soft X-ray energies. 

We performed standard data reduction with SAS~15 (\textsf{\small xmmsas\_20160201\_1833-15.0.0}) using up-to-date calibration files and following the analysis guidelines of each instrument\footnote{\texttt{http://xmm2.esac.esa.int/docs/documents}}.
We processed the EPIC exposures using the SAS meta tasks \textsf{\small epchain} and \textsf{\small emchain} and applied default corrections. 
For RGS, we used the SAS routine \textsf{\small rgsproc} to process the raw data files and create masks for the source and background regions.

Background flares were registered occasionally during the AO14 observations, usually lasting for less than a few kiloseconds. The percentages of good-time intervals (GTIs), filtering out periods of high background activity, are shown in Table~\ref{tab_dataexposure} for each scientific exposure and observation. Standard count rate thresholds were adopted for pn and MOS; for RGS, we used the background count rate on CCD 9 and applied a threshold of $0.1$\,s$^{-1}$ to filter the event lists with the SAS task \textsf{\small rgsfilter}. On average, data loss is small: of 2\% for MOS and RGS and 7\% for pn. The observation the worst affected by flares was that performed in August 2015, with a percentage of data loss between 9\% and 19\% depending on the camera. The total net exposures per camera are 303\,ks (MOS1), 312\,ks (MOS2), 314\,ks (RGS1), 312\,ks (RGS2), and 290\,ks (pn).

For the analysis of EPIC data, we filtered the event lists to exclude `bad' CCD pixels and columns, as well as to retain the pre-defined photon patterns with the highest quality energy calibration -- that are, single and double events for pn (pattern $\le4$) and single, double, triple, and quadruple for MOS (pattern $\le12$).
The source centroid and optimal extraction region, with typical sizes of 140'', were defined with the SAS task \textsf{\small eregionanalyse} in the $0.3-1.35$\,keV energy band for each EPIC camera and observation. Background circular regions of size $60''$ to $100''$ were defined away from the source, on the same CCD of the target whenever possible.

The detected source count rates, hardness ratios, the pile-up-relevant count rates per frame for each camera and observation, and other parameters based on a maximum likelihood fitting are listed in Table~\ref{tab_sourceparam}, with nominal $1\sigma$ statistical uncertainties. The parameters are determined with the SAS task \textsf{\small emldetect} on images created for each camera, observation, and energy band (only the combined EPIC results are shown; the X-ray emission of the source is compatible with the background level at energies above 2\,keV). For comparison, we also list the source parameters as determined from the 2012 EPIC exposure. 

Following the guidelines of \citet[][]{2015A&A...581A.104J}, we ensured that the pile-up levels at aimpoint were within tolerant guidelines for both EPIC detectors. Based on the source spectrum and on the number of counts per frame in each camera (listed in Table~\ref{tab_sourceparam} for direct comparison with Fig.~5 of \citealt{2015A&A...581A.104J}), we estimate that the percentage levels of spectral distortion and flux loss were around 1.5\% and 3.5\% for pn, and 0.3\% and 1\% for MOS. 

Overall, the source properties are consistent between epochs since 2012: potential discrepancies can be asserted to the cross-calibration uncertainties between the EPIC instruments and to different background levels. In Section~\ref{sec_spec} we investigate possible flux and spectral variations of the INS in detail.

We used the SAS task \textsf{\small eposcorr} to refine the astrometry by cross-correlating the list of EPIC X-ray source positions with those of catalogued near-infrared \citep[2MASS,][]{2006AJ....131.1163S}, optical \citep[GSC 2.3.2,][]{2008AJ....136..735L}, and X-ray \citep[\chan,][]{2010ApJS..189...37E} objects lying within $15'$ from \jsix. The results that yielded the least astrometric errors were obtained when cross-correlating the X-ray sources with a number of around 50 optical counterparts present in the field-of-view. Small positional offsets in right ascension and declination were consistently detected for all catalogues under study and are also shown in Table~\ref{tab_sourceparam}. The astrometrically corrected EPIC source positions in all four observations are consistent with each other. 

Finally, we verified the statistics of the EPIC lightcurves for general-trend variability. The lightcurves, binned into $600$\,s to $1200$\,s intervals, were corrected for bad pixels, dead-time, exposure, and background counts with the SAS task \textsf{\small epiclccorr}. All 2012/AO14 exposures are consistent with a constant flux.
\section{Analysis and results\label{sec_analysis}}
\subsection{Timing analysis\label{sec_timing}}
For the timing analysis we used a $Z^2_{\rm m}$ test \citep{1983A&A...128..245B} applied directly on the times-of-arrival of the pn and EPIC (pn+MOS) events to search for periodic signals. The times-of-arrival of the pn and MOS photons were converted from the local satellite to the solar system barycentric frame using the SAS task \textsf{\small barycen} and the astrometrically corrected source coordinates in each camera and observation (Table~\ref{tab_sourceparam}). 
\begin{figure*}
\begin{center}
\includegraphics*[width=0.79\textwidth]{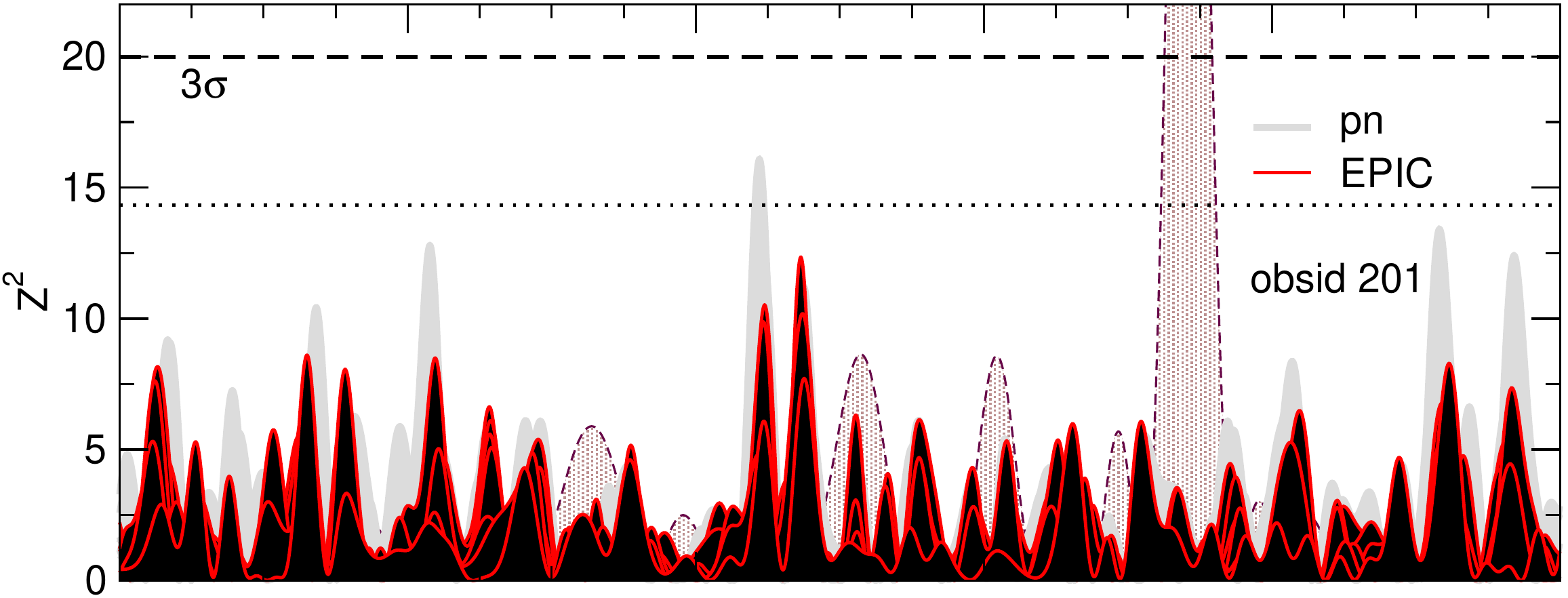}\\
\includegraphics*[width=0.79\textwidth]{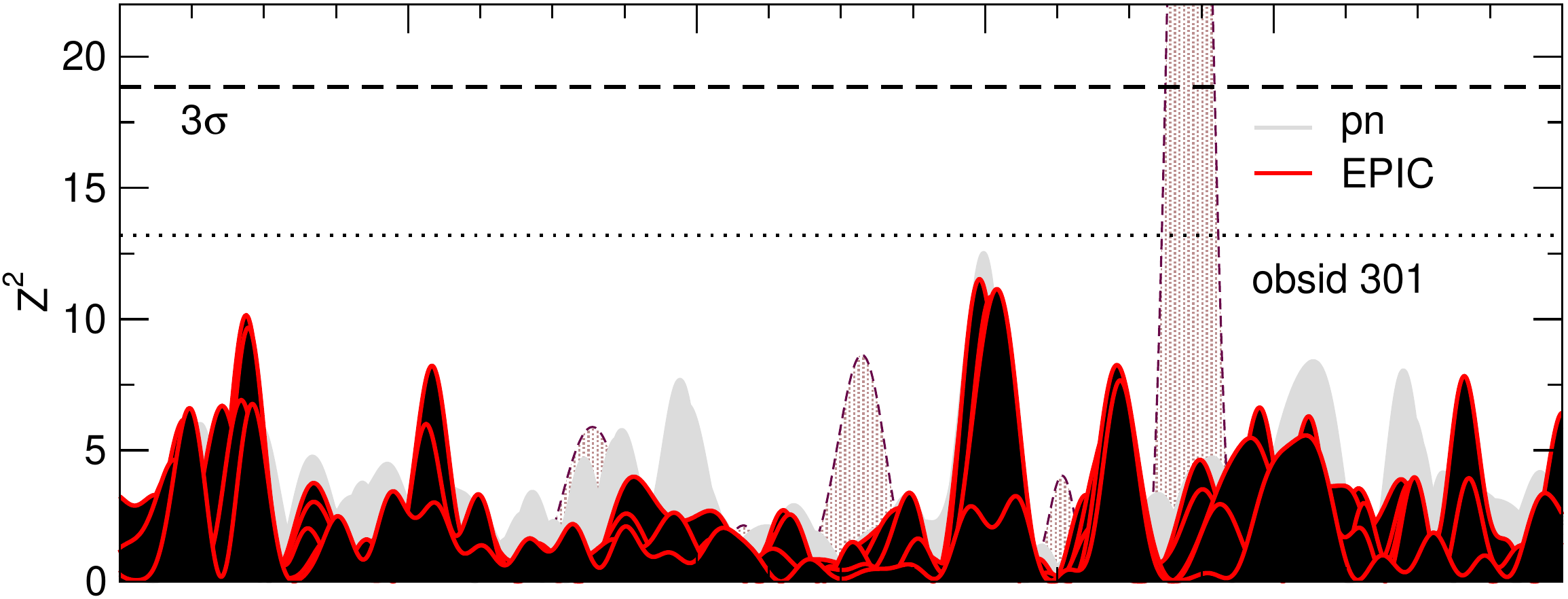}\\
\includegraphics*[width=0.79\textwidth]{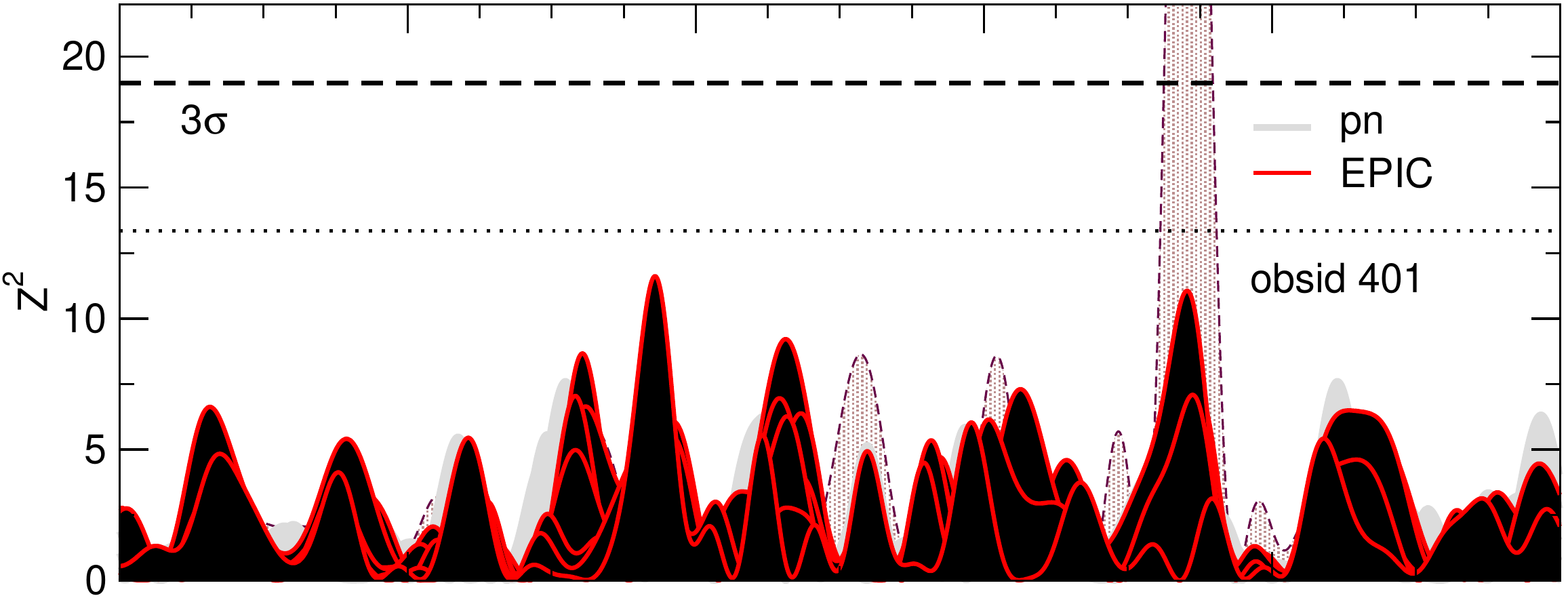}\\
\includegraphics*[width=0.79\textwidth]{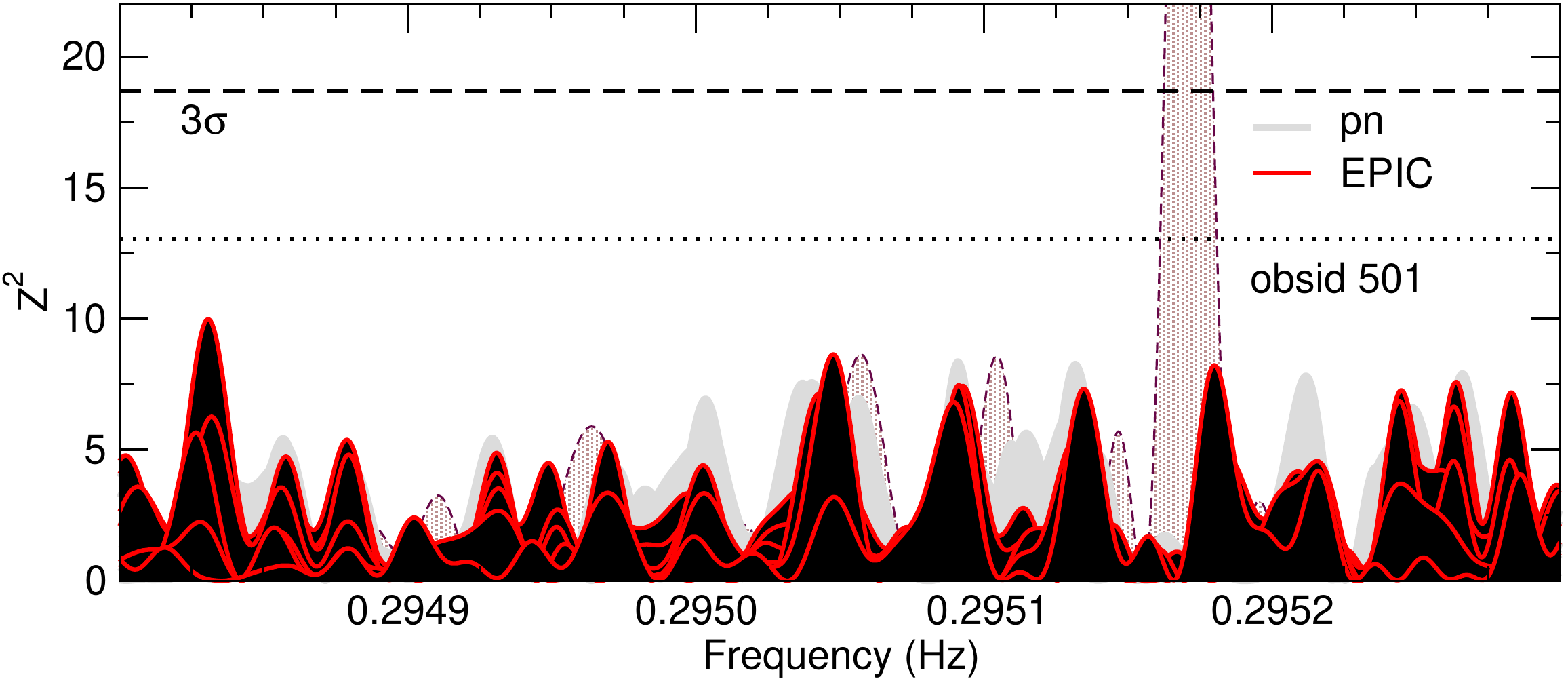}
\end{center}
\caption{$Z^2_1$ EPIC and pn searches around the 2012 periodicity. The frequency range is $\nu=0.2948-0.2953$\,Hz. The periodogram in the background (in filled brown colour and dashed outline) shows the 2012 result: $\nu_{2012}=0.2951709(14)$\,Hz and $Z^2_1(\nu_{2012})\sim50$ (\textsf{obsid} 0671620101, pn, $0.5-1.35$\,keV). The four plots show for each AO14 observation the results of tests conducted in seven different energy bands for an extraction region of $100''$ (see text). The dotted and dashed horizontal lines show the $2\sigma$ and $3\sigma$ confidence levels for the detection of modulations in each observation. \label{fig_Z2camera}}
\end{figure*}

Apart from the 2012 observation, the only other \xmm\ datasets suitable for timing analysis are a pn LW exposure performed in 2003 (0157360401, $\sim33$\,ks) and one performed in 2002 in timing (TI) mode (0073140501, $\sim30$\,ks); these observations were used to derive an estimate of the spin down rate of the source in combination with the 2012 dataset \citep{2014A&A...563A..50P}. The other archival pn observations of \jsix\ were either operated with the thick filter (hence reducing count rates by more than a factor of two) or severely affected by background flares. Likewise, the time resolution of 2.6\,s of the archival MOS observations of \jsix, all performed in FF mode, is not sufficient to detect the $3.39$\,s modulation. 

The re-analysis of the 2012 observation with SAS~15 and up-to-date calibration files yields similar results as those reported in \citet{2014A&A...563A..50P}. The only statistically significant modulation, at $\nu\equiv\nu_{2012}=0.2951709(14)$\,Hz, results when the search is restricted to source photons with energy above 0.5\,keV. We also verified that the periodic signal is always present at the same frequency within the errors, independently of the exact details of the processing of the raw event file (e.g.~included calibration files, SAS version, randomisation in energy within a PI channel, event filtering, or randomisation in time within the sampling detector time). The measured fluctuation of the power of the $Z^2_1$ statistic at $\nu_{2012}$, $Z^2_1(\nu_{2012})\sim35-49$ is consistent with that expected from a sinusoidal modulation of amplitude $p_{\rm f}=(4.1\pm0.9)$\% \citep[see, for example,][]{1999ApJ...511L..45P}. 

We first analysed each of the four AO14 observations individually. Taking into account the typical spin-down rate of the \msev\ INSs \citep[see, e.g.][and references therein]{2009ApJ...705..798K}, we looked for significant signals in a $5\times10^{-4}$\,Hz range around the 2012 frequency, adopting a resolution of $0.1$\,$\mu$Hz (oversampling factor of at least 10). The number of statistically independent trials in each search, which depends on the frequency range and on the total duration of the observation, is typically within 30 to 60. Assuming the usual scenario of magnetic dipole braking in vacuum, the $Z^2_{\rm m}$ tests allow for a maximum braking with respect to the 2012 signal of $|\dot{\nu}|\lesssim3.5\times10^{-12}$\,Hz\,s$^{-1}$, which corresponds to that exerted by a maximum dipolar magnetic field of $B_{\rm dip}\lesssim4\times10^{14}$\,G at the equator. 

Due to the energy-dependent nature of the 2012 signal \citep[see][for details]{2014A&A...563A..50P}, we carried out tests in various energy bands, also varying the size of the source extraction region and other parameters of the search (e.g~the included photon patterns and other details of the processing of the raw event files). We tested seven energy ranges in the soft ($0.15-0.5$\,keV, $0.2-0.5$\,keV, and $0.3-0.5$\,keV), hard ($0.5-1.35$\,keV), and total ($0.15-1.35$\,keV, $0.2-1.35$\,keV, and $0.3-1.35$\,keV) energy bands; source counts -- between $(0.18-5.5)\times10^5$ pn and $(0.4-7.2)\times10^5$ EPIC photons -- were extracted from circular regions of radius $10''$, $25''$, $50''$, $100''$ and $120''$ around the source position in each camera and epoch. Altogether, 1680 tests were conducted in the pn/EPIC datasets of the four observations. 
\begin{table}[t]
\caption{Upper limits on pulsations from the timing analysis
\label{tab_upperlim}}
\centering
\begin{tabular}{l c c c}
\hline\hline
 & EPIC ($0.01-0.56$\,Hz) & & pn ($<6.81$\,Hz) \\
\cline{2-4}
Band\tablefootmark{$\star$} & \multicolumn{3}{c}{Pulsed fraction} \\
 & \multicolumn{3}{c}{(\%, $3\sigma$ c.l.)} \\
\hline
soft & $1.36\pm0.05$ & & $1.60\pm0.10$ \\
hard & $2.01\pm0.07$ & & $2.78\pm0.16$ \\
total & $1.44\pm0.05$ & & $1.38\pm0.08$\\
\hline
\end{tabular}
\tablefoot{The $3\sigma$ upper limits are derived from the longest AO14 observation, \textsf{\small obsid} 0764460201. 
\tablefoottext{$\star$}{The soft, hard and total energy bands are, respectively: $0.2-0.5$\,keV, $0.5-1.35$\,keV and $0.3-1.35$\,keV, for EPIC (pn/MOS), and $0.15-0.5$\,keV, $0.5-1.65$\,keV, and $0.15-1.65$\,keV, for pn.}
}
\centering
\end{table}

The summary of results is in Figure~\ref{fig_Z2camera}. In each plot we show the $Z^2_1$ statistic as a function of trial frequency in each AO14 observation (pn/EPIC datasets), for different energy bands, and taking the extraction region of radius $100''$ as an illustrative example. The $Z^2$-test performed on the 2012 pn observation and respective peak at $Z^2_1(\nu_{2012})\sim50$ (out of scale) is shown in the background of the four plots for comparison ($0.5-1.35$\,keV). No significant power (above $3\sigma-4\sigma$) in the searched frequency range is consistently detected in these tests. The inclusion of higher harmonics (or different photon patterns and event filters) in the $Z^2_{\rm m}$ tests does not affect the results. Therefore, unless significant changes of pulsed fraction have taken place, $4\sigma$ upper limits\footnote{See, e.g.~\citet{1975ApJS...29..285G}, for the method to extract upper limits on the pulsed fraction.} of $1.33(6)$\%, $1.74(8)$\%, $1.84(8)$\%, and $1.80(8)$\% in the total energy band in each AO14 observation rule out the 2012 candidate period at a high confidence level.
\begin{table*}[t]
\caption{Results of the best-fit double blackbody model with a Gaussian absorption line (per observation and camera)
\label{tab_2bbg}}
\centering
\begin{tabular}{c c c r r r r r r r r r}
\hline\hline
\textsf{\small obsid} & $\chi^2_\nu$ & NHP & \multicolumn{1}{c}{$\nh$\tablefootmark{a}} & \multicolumn{1}{c}{$kT_1^\infty$} & \multicolumn{1}{c}{$kT_2^\infty$} & \multicolumn{1}{c}{$R_1^\infty$\tablefootmark{b}} & \multicolumn{1}{c}{$R_2^\infty$\tablefootmark{b}} & \multicolumn{1}{c}{$\epsilon$} & \multicolumn{1}{c}{$\sigma$} &  \multicolumn{1}{c}{$EW$} & \multicolumn{1}{c}{$f_{\rm X}$\tablefootmark{c}} \\
 & & (\%) & \multicolumn{1}{c}{} & \multicolumn{1}{c}{(eV)} & \multicolumn{1}{c}{(eV)} & \multicolumn{1}{c}{(km)} & \multicolumn{1}{c}{(km)}  &  \multicolumn{1}{c}{($3\sigma$, eV)} & \multicolumn{1}{c}{(eV)} & \multicolumn{1}{c}{($1\sigma$, eV)} & \multicolumn{1}{c}{}\\  
\hline
\multicolumn{12}{l}{pn}\\ 
2012  & $1.18$ & $26$ & $4.4_{-3}^{+1.6}$   & $64_{-4}^{+11}$ & $119.2_{-2.4}^{+2.6}$ & $13.8_{-5}^{+1.9}$ & $1.22_{-0.09}^{+0.10}$ & $<370$ & $119(10)$        & $<400$  & $6.37(5)$ \\ 
201   & $0.89$ & $88$ & $3.4_{-0.7}^{+1.3}$ & $64_{-4}^{+3}$  & $120.0_{-1.6}^{+1.7}$ & $12.2_{-2.0}^{+0.8}$ & $1.15(6)$ & $<360$ & $116_{-8}^{+4}$  & $<200$  & $6.73(3)$ \\  
301   & $1.04$ & $32$ & $5.4_{-1.0}^{+1.7}$ & $59(4)$         & $114.0_{-1.5}^{+1.7}$ & $18.4_{-3}^{+1.5}$ & $1.50_{-0.07}^{+0.08}$ & $<380$ & $114_{-11}^{+4}$ & $<220$  & $6.28(3)$ \\ 
401   & $0.96$ & $63$ & $4.2_{-1.2}^{+1.7}$ & $61_{-5}^{+6}$  & $117.6_{-1.9}^{+2.0}$ & $15.7_{-4}^{+1.6}$& $1.28_{-0.07}^{+0.08}$ & $<400$ & $111_{-9}^{+5}$  & $<250$  & $6.58(5)$ \\ 
501   & $0.88$ & $61$ & $5.4_{-1.3}^{+2.2}$ & $57(4)$         & $114.0(1.6)$          & $20.4_{-3}^{+2.1}$& $1.55(6)$ & $<370$ & $109_{-12}^{+5}$ & $<240$  & $6.46(4)$ \\
\hline
\multicolumn{12}{l}{MOS\tablefootmark{d}}\\
2012  & $1.02$ & $41$ & $1.5_{-1.4}^{+1.6}$ & $74_{-10}^{+9}$ & $127(3)$   & $6.3_{-2.4}^{+0.8}$ & $0.84_{-0.06}^{+0.07}$ & $<430$  & $95_{-16}^{+20}$  & $<90$  & $6.58(6)$ \\ 
201  & $1.09$ & $19$ & $1.8_{-0.7}^{+1.2}$ & $70_{-5}^{+4}$  & $125.6(2.2)$  & $8.1_{-1.6}^{+0.7}$ & $0.88(5)$ & $<350$  & $119_{-8}^{+3}$   & $<190$  & $6.70(4)$ \\ 
301  & $0.95$ & $66$ & $2.0_{-1.4}^{+1.6}$ & $77_{-8}^{+9}$  & $131(4)$    & $5.8_{-1.9}^{+0.5}$ & $0.73_{-0.06}^{+0.07}$ & $<450$  & $91_{-15}^{+17}$  & $<75$  & $6.29(4)$ \\ 
401  & $1.05$ & $31$ & $2.1_{-1.0}^{+1.4}$ & $69(5)$         & $126(3)$    & $9.0_{-1.6}^{+0.9}$ & $0.85(6)$ & $<480$  & $120_{-13}^{+4}$  & $<220$  & $6.58(5)$ \\ 
501  & $1.06$ & $22$ & $3.6_{-2.3}^{+2.2}$ & $43_{-5}^{+8}$ & $118.3_{-1.6}^{+1.7}$ & $>20$ & $1.26(6)$ & $520(30)$ & $121_{-12}^{+13}$ & $<160$ & $7.01(6)$\\ 
\hline
\multicolumn{12}{l}{EPIC\tablefootmark{e}}\\
2012 & $1.15$ & $2.2$ & $3.9_{-0.9}^{+1.0}$ & $64.2_{-2.8}^{+6}$ & $120.5(1.8)$ & $13.3_{-2.9}^{+1.0}$ & $1.16(5)$ & $<390$ & $117_{-16}^{+3}$ & $<200$ & $6.40(3)$ \\ 
201  & $1.00$ & $48$ & $2.4_{-0.9}^{+0.5}$ & $67.3_{-1.7}^{+3}$ & $122.3(1.2)$ & $10.4_{-1.2}^{+0.5}$ & $1.07(3)$ & $<340$ & $118.6_{-4}^{+2.5}$ & $<150$ & $6.885(25)$ \\ 
301  & $1.02$ & $37$ & $3.2_{-1.3}^{+1.4}$ & $68_{-5}^{+6}$ & $120.7_{-1.6}^{+1.7}$ & $9.5_{-2.3}^{+0.6}$ & $1.12(5)$ & $<420$ & $103_{-13}^{+15}$ & $<150$ & $6.45(3)$ \\ 
401  & $1.01$ & $41$ & $2.9_{-0.7}^{+1.1}$ & $65_{-3}^{+4}$ & $121.6(1.8)$ & $11.3_{-1.9}^{+0.8}$ & $1.06(5)$ & $<370$ & $116_{-10}^{+3}$ & $<160$ & $6.77(3)$ \\
501  & $1.08$ & $14$ & $3.0_{-0.6}^{+1.2}$ & $66.4_{-6}^{+1.2}$ & $123.6(1.8)$ & $10.9_{-2.1}^{+0.7}$ & $1.01_{-0.04}^{+0.05}$ & $<350$ & $118_{-7}^{+3}$ & $<150$ & $6.82(3)$ \\ 
\hline
\multicolumn{12}{l}{Multi-epoch fits\tablefootmark{f}}\\
pn & $1.02$ & $29$ & $4.5_{-0.6}^{+0.5}$ & $60.9_{-1.5}^{+1.7}$ & $117.0(8)$ & $16.2_{-1.3}^{+0.6}$ & $1.34(3)$ & $<320$ & $114.6_{-2.6}^{+2.0}$ & $100^{+50}_{-60}$ &  $6.468(16)$ \\ 
MOS  & $1.11$ & $<1$ & $2.5_{-0.4}^{+0.7}$ & $68.4_{-3}^{+2.0}$ & $125.7(1.2)$ & $9.7_{-1.1}^{+0.4}$ & $0.90(3)$ & $<340$ & $118.9_{-5}^{+1.6}$ & $100^{+50}_{-40}$ & $6.877(22)$ \\
EPIC & $1.12$ & $<1$ & $3.1_{-0.3}^{+0.5}$ & $65.8_{-2.2}^{+1.0}$ & $121.7(7)$ & $11.8_{-0.9}^{+0.3}$ & $1.101(20)$ & $<320$ & $118.1_{-2.8}^{+1.3}$ & $90^{+40}_{-10}$ & $6.663(12)$ \\ 
\hline
\hline
\multicolumn{4}{l}{Multi-epoch fits\tablefootmark{g}} & \multicolumn{2}{c}{$kT^\infty$ (eV)} & \multicolumn{2}{c}{$R^\infty$ (km)} & & & & \\
\cline{5-6}\cline{7-8} 
pn & $3.8$ & $\ll1$ & $2.4^\star$ & \multicolumn{2}{c}{$88.16_{-0.19}^{+0.18}$} & \multicolumn{2}{c}{$5.16_{-0.04}^{+0.03}$} & $<300$ & $170.7(8)$ & $135_{-19}^{+6}$ & $6.077(7)$ \\
MOS & $2.5$ & $\ll1$ & $2.4^\star$ & \multicolumn{2}{c}{$91.12_{-0.29}^{+0.3}$} & \multicolumn{2}{c}{$4.39_{-0.05}^{+0.04}$} & $<305$ & $180.8(1.5)$ & $117_{-14}^{+26}$ & $5.678(10)$ \\
EPIC & $5.5$ & $\ll1$ & $2.4^\star$ & \multicolumn{2}{c}{$89.18_{-0.16}^{+0.14}$} & \multicolumn{2}{c}{$4.94\pm0.03$} & $<300$ & $174.1(6)$ & $112.9_{-1.1}^{+19}$ & $5.928(5)$ \\
\hline
\end{tabular}
\tablefoot{Errors are $1\sigma$ confidence levels. The model fitted to the data in XSPEC is \textsf{\small tbabs(bbody+bbody-gauss)}. The degrees of freedom (d.o.f.) of each fit are within 182 and 202 (pn spectra), 179 and 197 (MOS), 368 and 406 (EPIC). The d.o.f.~of the multi-epoch double blackbody fits are 992, 1736, and 1200 (for pn, MOS, and EPIC, respectively).
\tablefoottext{a}{The column density is in units of $10^{20}$\,cm$^{-2}$.}
\tablefoottext{b}{The radiation radius at infinity for each component is computed from the derived blackbody luminosity for a source at a distance of $d\equiv d_{300}=300$\,pc.}
\tablefoottext{c}{The observed model flux is in units of $10^{-12}$\,erg\,s$^{-1}$\,cm$^{-2}$ in energy band $0.2-12$\,keV.}
\tablefoottext{d}{Simple fit of combined MOS1 and MOS2 spectra (per observation).}
\tablefoottext{e}{Simultaneous fit of pn and combined MOS spectra (per observation).}
\tablefoottext{f}{Simultaneous fit per instrument (5 pn, 10 MOS1/2, and 6 pn and stacked MOS spectra).}
\tablefoottext{g}{Results of simultaneous fits of a single temperature model, \textsf{\small tbabs(bbody-gauss)}.}
} 
\end{table*}

We next searched the new data, especially the longest and most sensitive `201' observation, for other significant modulations in the full frequency range allowed by the resolution of the EPIC cameras (`blind searches'). To this end, the times-of-arrival of the pn and MOS events ($\sim2.5\times10^5$ counts) were analysed together in the $\nu=0.01-0.56$\,Hz frequency range; for pn, timing searches were extended to higher frequencies, up to $\sim6.8$\,Hz\footnote{The maximum frequency is determined by the Nyquist limit, given the time resolution of the detector.}. The adopted frequency step was $\Delta\nu=2.5-5$\,$\mu$Hz (oversampling factor of 3) and the number of independent trials were $(4-8)\times10^5$ and $(3-6)\times10^4$ in the pn and EPIC searches, respectively. A total of seven `narrow' (with widths between $100$\,eV and $600$\,eV) and three `wide' ($600-850$\,eV) energy bands, defined within the $0.15-1.35$\,keV range according to the source's signal-to-noise ratio, were defined for these searches. We tested three different sizes of source extraction regions ($25''$, $50''$, and $120''$) and included all valid photon patterns in the EPIC searches. For pn, we restricted the event lists to include only single and double photon patterns. In total, 240 EPIC and pn blind searches were performed in the observations of the large programme.

No significant periodic signal resulted from the analysis. In Table~\ref{tab_upperlim} we list the most constraining $3\sigma$ upper limits from the EPIC and pn searches for three wide energy band intervals (soft, hard, and total; see the table caption for details). 
\subsection{Spectral analysis\label{sec_spec}}
\subsubsection{EPIC data\label{sec_epicspec}}
\begin{figure*}
\begin{center}
\includegraphics*[width=0.495\textwidth]{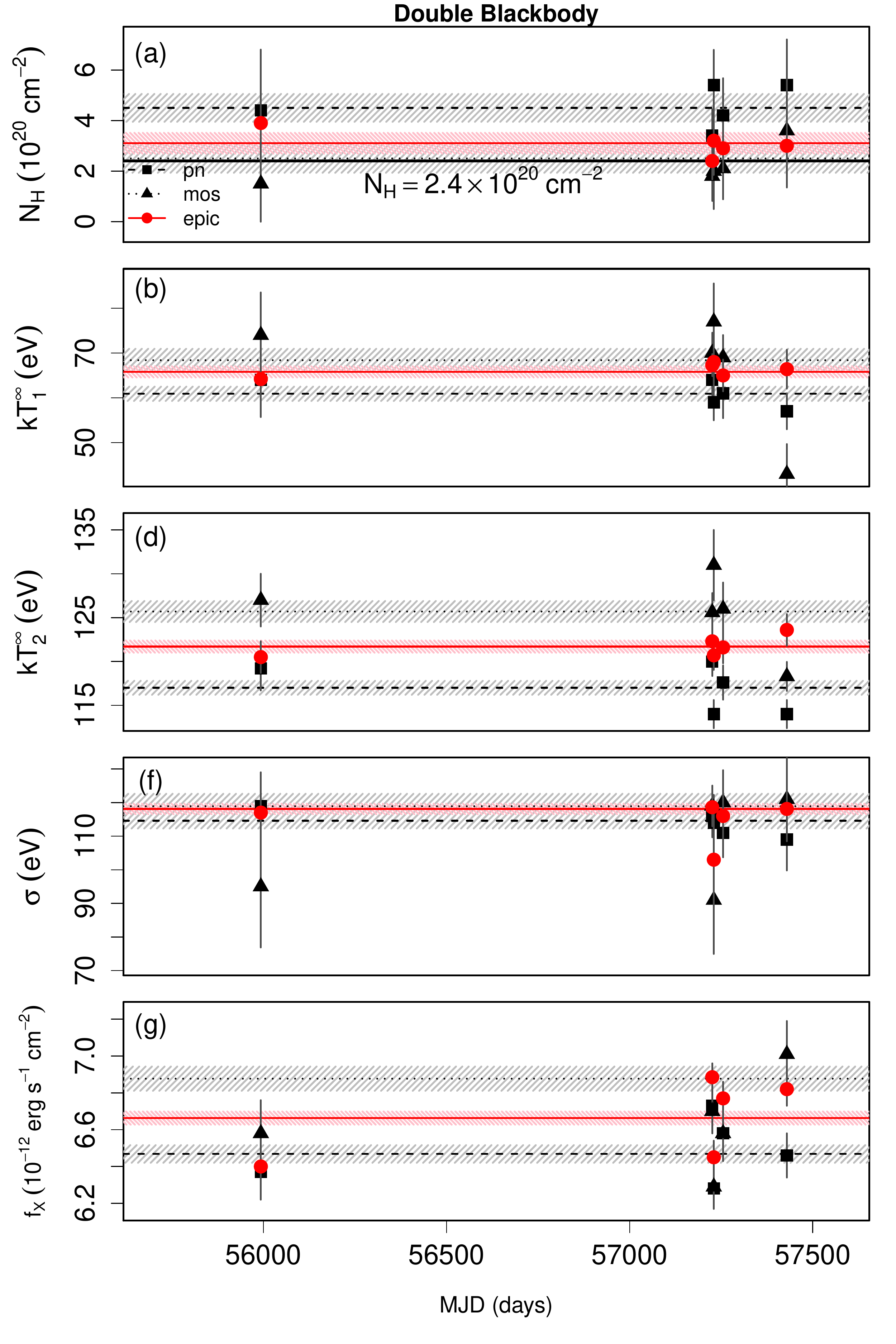}
\includegraphics*[width=0.495\textwidth]{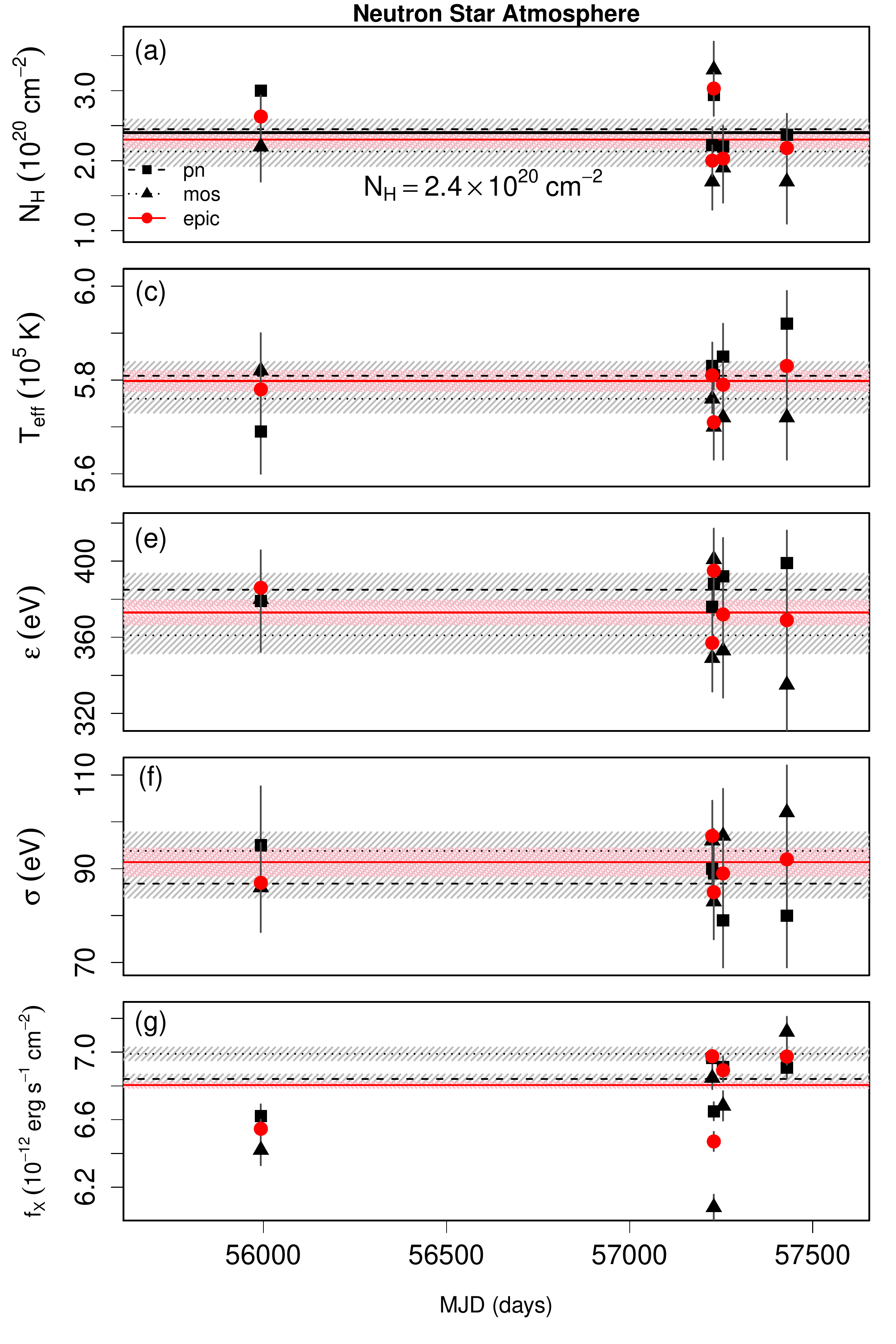}
\end{center}
\caption{Source best-fit parameters as a function of MJD (EPIC analysis; see the text and Tables~\ref{tab_2bbg} and \ref{tab_nsag}, for details). The model fit to the data is that of absorbed double blackbody (left) and fully ionized hydrogen neutron star atmosphere ($B=10^{13}$\,G, $M=1.4$\,M$_\odot$, and $R=10$\,km; right), modified by a broad Gaussian absorption line. The plots show, in each instrument and observation (see legend): (a) the column density, the temperature of the (b) cold and (d) hot blackbody components, (c) the effective temperature of the neutron star (unredshifted), (e) the central energy of the absorption line, (f) its Gaussian sigma, and (g) the observed model flux in the $0.2-12$\,keV energy band. The total Galactic $\nh$ value is shown by the solid black line in plots (a). Horizontal lines show the results of the simultaneous fit of spectra over the five pointings, with $1\sigma$ confidence levels comprised by the shaded areas.\label{fig_2bbNSAgpar}}
\end{figure*}

The analysis of the EPIC data is based on source and background spectra extracted from regions as described in Section~\ref{sec_data}, together with the respective response matrices and ancillary files created for each of the EPIC cameras and observation. In accordance to the guidelines and calibration status of the instruments, we restricted the spectral analysis to GTI-filtered photons with energies between 0.3\,keV and 1.35\,keV (beyond which the source signal-to-noise ratio becomes insignificant).
Including the 2012 observation, the analysed dataset comprises 15 spectra and over $1.2\times10^6$ counts ($0.3-1.35$\,keV), of which 1.8\% can be ascribed to the background. 

The pile-up level is negligible in the MOS exposures (Section~\ref{sec_data}); for pn, we applied a correction in the redistribution matrix files with the SAS task \textsf{\small rmfgen} to minimise flux loss and spectral distortion\footnote{\texttt{https://www.cosmos.esa.int/web/xmm-newton/\\sas-thread-epatplot}}. The correction is calculated directly from the frequency and spectrum of the incoming photons and has the advantage of keeping events from the central PSF area in the spectral analysis, which would have to be otherwise discarded. We verified that the results using this approach agree well with those when the spectra are extracted from regions with a $10''-15''$ excised core, while avoiding up to $30\%$ of data loss in the pn exposures.

The energy channels of each spectrum, which are by construction 5\,eV wide, were regrouped to avoid a low ($<30$) number of counts per spectral bin. Due to the brightness of the source and the good statistics of each individual spectrum this has an effect only at the high energy side of the analysis, where the source signal becomes dominated by the background. At lower energies ($\lesssim700$\,eV), the spectrum oversamples the instrument resolution of the EPIC cameras by a factor of up to 20. We ensured nonetheless that oversampling did not influence the results of spectral fitting -- specifically, we checked for consistency where oversampling was kept within a maximum factor of 3.

To fit the spectra we used XSPEC~12.9.0n \citep{1996ASPC..101...17A}. Unless otherwise noted, the fit parameters were allowed to vary freely within reasonable ranges. The photoelectric absorption model and elemental abundances of \citet[][\textsf{\small tbabs} in \textsf{\small XSPEC}]{2000ApJ...542..914W} were adopted to account for the interstellar material in the line-of-sight. Due to the low absorption towards \jsix, the choice of abundance table and cross-section model does not significantly impact the results of the spectral fitting. 

\begin{table*}[t]
\caption{Results of the best-fit fully ionized neutron star atmosphere model with a Gaussian absorption line (per observation and camera)
\label{tab_nsag}}
\centering
\begin{tabular}{c c c c r r c r r r r}
\hline\hline
\textsf{\small obsid} & $\chi^2_\nu$ & d.o.f & NHP & \multicolumn{1}{c}{$\nh$} & \multicolumn{1}{c}{$T_{\rm eff}$} & \multicolumn{1}{c}{$d$} & \multicolumn{1}{c}{$\epsilon$} & \multicolumn{1}{c}{$\sigma$} &  \multicolumn{1}{c}{$EW$} & \multicolumn{1}{c}{$f_{\rm X}$\tablefootmark{a}} \\
 & & & (\%) & \multicolumn{1}{c}{($10^{20}$\,cm$^{-2}$)} & \multicolumn{1}{c}{($10^5$\,K)} & \multicolumn{1}{c}{(pc)} & \multicolumn{1}{c}{(eV)} & \multicolumn{1}{c}{(eV)} & \multicolumn{1}{c}{(eV)} & \multicolumn{1}{c}{}\\  
\hline
\multicolumn{11}{l}{pn}\\ 
2012 & $1.19$ & 184 & $4$ & $3.0(4)$ & $5.69_{-0.08}^{+0.09}$ & $116(8)$ & $380_{-30}^{+23}$ & $95_{-11}^{+14}$ & $70_{-40}^{+3}$ & $6.622(23)$\\
201 & $0.90$ & 204 & $84$ & $2.22_{-0.26}^{+0.25}$ & $5.83(5)$ & $130(5)$ & $376_{-18}^{+15}$ & $90_{-7}^{+8}$ & $55_{-4}^{+5}$ & $6.965(16)$ \\
301 & $1.07$ & 195 & $25$ & $2.9(3)$ & $5.68(7)$ & $116(6)$ & $388_{-22}^{+17}$ & $88_{-9}^{+10}$ & $66_{-40}^{+1.0}$ & $6.650(18)$ \\
401 & $0.97$ & 195 & $62$ & $2.2(3)$ & $5.85(7)$ & $134_{-7}^{+6}$ & $392_{-23}^{+17}$ & $79_{-9}^{+11}$ & $70_{-30}^{+1}$ & $6.912(21)$ \\
501 & $1.01$ & 196 & $45$ & $2.37_{-0.3}^{+0.29}$ & $5.92(7)$ & $137(7)$ & $399_{-19}^{+15}$ & $80^{+8}_{-10}$ & $50_{-6}^{+4}$ & $6.909(21)$ \\
\hline
\multicolumn{11}{l}{MOS\tablefootmark{b}}\\
2012 & $1.05$ & 181 & $30$ & $2.2(5)$ & $5.82_{-0.07}^{+0.08}$ & $134_{-9}^{+8}$ & $380_{-25}^{+19}$ & $86_{-9}^{+10}$ & $76_{-30}^{+1.0}$ & $6.42(3)$ \\
201 & $1.11$ & 199 & $14$ & $1.81(4)$ & $5.79_{-0.06}^{+0.05}$ & $131_{-6}^{+5}$ & $338_{-25}^{+20}$ & $102_{-8}^{+10}$ & $65_{-25}^{+4}$ & $6.741_{-0.020}^{+0.022}$ \\ 
301 & $0.98$ & 187 & $59$ & $3.3(4)$ & $5.70(7)$ & $120_{-7}^{+6}$ & $401_{-18}^{+14}$ & $83_{-7}^{+9}$ & $75_{-30}^{+2.0}$ & $6.081(25)$ \\
401 & $1.08$ & 185 & $23$ & $1.9(5)$ & $5.72(9)$ & $128_{-9}^{+7}$ & $353_{-28}^{+21}$ & $97_{-9}^{+11}$ & $75_{-40}^{+1.0}$ & $6.682(29)$ \\
501 & $1.12$ & 188 & $12$ & $1.8(6)$ & $5.72_{-0.09}^{+0.07}$ & $124_{-8}^{+7}$ & $335_{-30}^{+24}$ & $102_{-9}^{+12}$ & $76_{-40}^{+1.0}$ & $7.02(3)$ \\
\hline
\multicolumn{11}{l}{EPIC\tablefootmark{c}}\\
2012 & $1.17$ & 370 & $1.3$ & $2.63_{-0.3}^{+0.29}$ & $5.78(6)$ & $124(5)$ & $386_{-18}^{+15}$ & $87_{-7}^{+8}$ & $55_{-7}^{+3}$ & $6.546(18)$\\
201 & $1.02$ & 408 & $36$ & $2.00_{-0.21}^{+0.22}$ & $5.81(4)$ & $129_{-4}^{+3}$ & $357_{-15}^{+12}$ & $97_{-5}^{+6}$ & $60_{-3}^{+4}$ & $6.976(11)$ \\
301 & $1.03$ & 387 & $31$ & $3.03(23)$ & $5.71(5)$ & $129(6)$ & $395_{-14}^{+11}$ & $85_{-6}^{+7}$ & $55_{-2.0}^{+4}$ & $6.471(19)$\\
401 & $1.03$ & 385 & $34$ & $2.03(28)$ & $5.79(5)$ & $129(5)$ & $372_{-18}^{+14}$ & $89_{-7}^{+8}$ & $60_{-3}^{+6}$ & $6.893(22)$\\
501 & $1.10$ & 389 & $8$ & $2.18(28)$ & $5.83(5)$ & $128_{-6}^{+4}$ & $369_{-18}^{+14}$ & $92_{-6}^{+8}$ & $60_{-3}^{+7}$ & $6.974(16)$\\
\hline
\multicolumn{11}{l}{Multi-epoch fits\tablefootmark{d}}\\
pn & $1.04$ & 994 & $17$ & $2.45(14)$ & $5.81(3)$ & $128.2_{-3}^{+2.6}$ & $385_{-9}^{+8}$ & $87(4)$ & $54_{-2.0}^{+5}$ & $6.842(9)$ \\ 
MOS & $1.13$ & 1738 & $<1$ & $2.13_{-0.19}^{+0.21}$ & $5.763_{-0.04}^{+0.026}$ & $126.7_{-4}^{+2.6}$ & $361_{-11}^{+9}$ & $93(4)$ & $64_{-2.0}^{+4} $ & $7.016(13)$ \\ 
EPIC & $1.15$ & 1202 & $<1$ & $2.30_{-0.11}^{+0.12}$ & $5.798_{-0.026}^{+0.018}$ & $127.3_{-2.4}^{+1.8}$ & $373_{-7}^{+6}$ & $91.2_{-2.8}^{+3}$ & $57.7_{-1.4}^{+2.1}$ & $6.805(6)$ \\ 
\hline
\end{tabular}
\tablefoot{Errors are $1\sigma$ confidence levels. The model fitted to the data in XSPEC is \textsf{\small tbabs(nsa-gauss)}, assuming a magnetic field intensity of $B=10^{13}$\,G and canonical neutron star mass and radius. The effective temperature is given at the source's rest frame (unredshifted). 
\tablefoottext{a}{The observed model flux is in units of $10^{-12}$\,erg\,s$^{-1}$\,cm$^{-2}$ in energy band $0.2-12$\,keV.}
\tablefoottext{b}{Simple fit of combined MOS1 and MOS2 spectra (per observation).}
\tablefoottext{c}{Simultaneous fit of pn and combined MOS spectra (per observation).}
\tablefoottext{d}{Simultaneous fit per instrument (5 pn, 10 MOS1/2, and 6 pn and stacked MOS spectra).}
} 
\end{table*}

The 15 spectra were first fit individually to check the agreement between the instruments and epochs. The exercise showed the expected few percent cross-calibration uncertainty between the EPIC detectors \citep{2014A&A...564A..75R}. Although variations from pointing to pointing for a given instrument are formally significant with respect to a constant value, the relative error is still smaller than the absolute discrepancies between the cameras within a given epoch. To account for this uncertainty, we allowed for a renormalisation factor in XSPEC and fitted the spectra of the pn and MOS cameras simultaneously. 
We checked that the results of the simultaneous fits were consistent with the weighted means of the individual measurements. 

We then checked if consistent results are obtained when fitting a single `stacked' spectrum, which converges to the best parameter values much faster in XSPEC than the simultaneous fits. The stacked spectra are produced with the SAS task \textsf{\small epicspeccombine}, taking into account the responses, background, and ancillary files of the individual exposures. For pn, the stacking approach leads to inconsistent results that do not match the corresponding weighted mean values for the camera within the errors, nor the results of the simultaneous fits\footnote{See {\small \texttt{https://www.cosmos.esa.int/web/xmm-newton/\\sas-thread-epic-merging\#cav}}, for details.}. On the other hand, the results of the combined MOS spectra agree well with those of the simultaneous fits. We adopted hereafter a stacked spectrum only for MOS to avoid introducing biased results in the spectral analysis.
\begin{table*}[t]
\caption{Results of the best-fit partially ionized neutron star atmosphere model with a Gaussian absorption line (per observation and camera)
\label{tab_nsmaxg}}
\centering
\begin{tabular}{c c c c r r c c r r r r}
\hline\hline
\textsf{\small obsid} & $\chi^2_\nu$ & d.o.f. & NHP & \multicolumn{1}{c}{$\nh$\tablefootmark{a}} & \multicolumn{1}{c}{$T_{\rm eff}$} & \multicolumn{1}{c}{$d$} & \multicolumn{1}{c}{$R_{\rm em}$} & \multicolumn{1}{c}{$\epsilon$} & \multicolumn{1}{c}{$\sigma$} &  \multicolumn{1}{c}{$EW$} & \multicolumn{1}{c}{$f_{\rm X}$\tablefootmark{a}} \\
 & & & (\%) & \multicolumn{1}{c}{($10^{20}$\,cm$^{-2}$)} & \multicolumn{1}{c}{($10^5$\,K)} & \multicolumn{1}{c}{($3\sigma$, pc)} & \multicolumn{1}{c}{(km)} & \multicolumn{1}{c}{(eV)} & \multicolumn{1}{c}{(eV)} & \multicolumn{1}{c}{(eV)} & \multicolumn{1}{c}{}\\  
\hline
\multicolumn{12}{l}{pn}\\ 
2012 & $1.18$ & 185 & $5$ & $4.07(17)$ & $6.17_{-0.06}^{+0.11}$ & $<107$ & $9.05_{-0.07}^{+0.4}$ & $382_{-16}^{+9}$ & $89(5)$ & $52_{-12}^{+3}$ & $6.581(23)$\\
201 & $0.88$ & 205 & $89$ & $4.12_{-0.21}^{+0.25}$ &$5.85_{-0.09}^{+0.07}$ & $<120$ & $10.3_{-0.5}^{+0.8}$ & $362_{-17}^{+12}$ & $92_{-5}^{+7}$ & $61.7_{-1.5}^{+3}$ & $6.708_{-0.012}^{+0.014}$ \\ 
301 & $1.07$ & 196 & $25$ & $4.06_{-0.11}^{+0.10}$ & $6.16_{-0.08}^{+0.04}$ & $<105$ & $9.0_{-0.7}^{+0.5}$ & $390(6)$ & $84(4)$ & $50_{-10}^{+1.0}$ & $6.589(18)$ \\ 
401 & $0.96$ & 196 & $65$ & $3.32_{-0.22}^{+0.4}$ & $6.33_{-0.14}^{+0.3}$ & $<110$ & $8.7_{-0.8}^{+0.6}$ & $387_{-22}^{+11}$ & $80_{-7}^{+11}$ & $50(5)$ & $6.828_{-0.20}^{+0.19}$ \\
501 & $1.00$ & 197 & $50$ & $3.58_{-0.21}^{+0.4}$ & $6.34_{-0.14}^{+0.20}$ & $113(6)$ & $9.0_{-0.7}^{+0.6}$ & $394_{-20}^{+11}$ & $81_{-6}^{+10}$ & $48_{-6}^{+2}$ & $6.409_{-0.018}^{+0.016}$ \\
\hline 
\multicolumn{12}{l}{MOS\tablefootmark{b}}\\
2012 & $1.06$ & 182 & $29$ & $3.4(4)$ & $6.30_{-0.15}^{+0.5}$ & $<117$ & $8.7_{-1.2}^{+0.6}$ & $373_{-17}^{+15}$ & $87(8)$ & $63_{-5}^{+9}$ & $6.311_{-0.03}^{+0.029}$ \\
201 & $1.10$ & 200 & $16$ & $3.12(29)$ & $6.16_{-0.14}^{+0.13}$ & $<110$ & $9.0_{-0.7}^{+0.9}$ & $325_{-19}^{+15}$ & $104(6)$ & $68_{-4}^{+3}$ & $6.608_{-0.020}^{+0.021}$ \\
301 & $0.98$ & 188 & $54$ & $4.30_{-0.17}^{+0.13}$ & $6.21_{-0.08}^{+0.12}$ & $<105$ & $8.82_{-0.7}^{+0.28}$ & $396(7)$ & $82(4)$ & $56_{-3}^{+4}$ & $6.072(25)$ \\
401 & $1.07$ & 186 & $25$ & $3.09_{-0.3}^{+0.21}$ & $6.22_{-0.18}^{+0.17}$ & $<110$ & $8.6_{-0.8}^{+0.7}$ & $344_{-13}^{+16}$ & $99_{-7}^{+6}$ & $66_{-9}^{+7}$ & $6.555_{-0.028}^{+0.027}$ \\ 
501 & $1.12$ & 189 & $12$ & $2.84_{-0.27}^{+0.20}$ & $6.25_{-0.16}^{+0.17}$ & $<110$ & $8.6_{-0.9}^{+0.6}$ & $322_{-12}^{+14}$ & $105_{-6}^{+5}$ & $83_{-12}^{+1.0}$ & $6.602(27)$ \\ 
\hline
\multicolumn{12}{l}{EPIC\tablefootmark{c}}\\
2012 & $1.16$ & 371 & $1.8$ & $4.11_{-0.18}^{+0.12}$ & $6.02_{-0.08}^{+0.04}$ & $<110$ & $9.5_{-0.5}^{+0.4}$ & $375_{-7}^{+10}$ & $89_{-5}^{+4}$ & $58_{-8}^{+3}$ & $6.419(18)$ \\
201  & $1.01$ & 409 & $44$ & $3.50(16)$ & $6.06_{-0.09}^{+0.06}$ & $<110$ & $9.38_{-0.28}^{+0.4}$ & $343_{-11}^{+10}$ & $99_{-4}^{+5}$ & $63_{-2.2}^{+3}$ & $6.391(10)$ \\
301  & $1.03$ & 388 & $31$ & $4.11(10)$ & $6.18(7)$ & $<105$ & $8.95_{-0.4}^{+0.19}$ & $392(4)$ & $83.2(2.5)$ & $49_{-6}^{+1.0}$ & $6.430_{-0.013}^{+0.015}$ \\
401 & $1.02$ & 386 & $39$ & $3.23_{-0.23}^{+0.19}$ & $6.27_{-0.14}^{+0.12}$ & $<110$ & $8.7(6)$ & $364_{-12}^{+11}$ & $91(6)$ & $58(4)$ & $6.769(16)$ \\
501 & $1.10$ & 390 & $10$ & $3.55_{-0.22}^{+0.18}$ & $6.17_{-0.13}^{+0.08}$ & $<110$ & $9.2_{-0.6}^{+0.5}$ & $357_{-13}^{+12}$ & $95_{-5}^{+6}$ & $62(3)$ & $6.424_{-0.015}^{+0.013}$ \\
\hline
\multicolumn{12}{l}{Multi-epoch fits\tablefootmark{d}}\\
pn & $1.03$ & 995 & $26$ & $4.85(14)$ & $5.56_{-0.04}^{+0.03}$ & $106_{-4}^{+6}$ & $11.9_{-0.4}^{+0.6}$ & $362_{-9}^{+8}$ & $92_{-3}^{+4}$ & $62.5_{-1.5}^{+2.5}$ & $6.519_{-0.007}^{+0.008}$ \\ 
MOS & $1.13$ & 1739 & $<1$ & $3.31_{-0.13}^{+0.07}$ & $6.24_{-0.08}^{+0.05}$ & $<105$ & $8.66_{-0.08}^{+0.3}$ & $353_{-4}^{+6}$ & $95.0_{-2.8}^{+1.9}$ & $64.1_{-5}^{+1.0}$ & $6.875_{-0.011}^{+0.013}$ \\ 
EPIC & $1.14$ & 1203 & $<1$ & $3.82(10)$ & $6.04_{-0.05}^{+0.03}$ & $<105$ & $9.41_{-0.23}^{+0.25}$ & $361_{-4}^{+5}$ & $93.6_{-2.1}^{+2.0}$ & $60.5_{-2.8}^{+0.6}$ & $6.227_{-0.006}^{+0.007}$ \\ 
\hline
\end{tabular}
\tablefoot{Errors are $1\sigma$ confidence levels. The model fitted to the data in XSPEC is \textsf{\small tbabs(nsmaxg-gauss)}, assuming a magnetic field intensity of $B=10^{13}$\,G and a $1.4$\,M$_\odot$ neutron star. The effective temperature and radius are given at the source's rest frame (unredshifted). 
\tablefoottext{$\star$}{Parameter held fixed during spectral fitting.}
\tablefoottext{a}{The observed model flux is in units of $10^{-12}$\,erg\,s$^{-1}$\,cm$^{-2}$ in energy band $0.2-12$\,keV.}
\tablefoottext{b}{Simple fit of combined MOS1 and MOS2 spectra (per observation).}
\tablefoottext{c}{Simultaneous fit of pn and combined MOS spectra (per observation).}
\tablefoottext{d}{Simultaneous fit per instrument (5 pn, 10 MOS1/2, and 6 pn and stacked MOS spectra).}
} 
\end{table*}

Next, we proceeded at finding a model which closely describes the X-ray spectral energy distribution of the source (see \citealt{2014A&A...563A..50P}, for details). We list in Table~\ref{tab_2bbg} the results of the fit of a double blackbody model\footnote{For comparison, we also list in Table~\ref{tab_2bbg} for the multi-epoch fits the results of a single temperature blackbody model with a Gaussian absorption line. For this model, the fit quality is generally poor ($\chi^2_\nu\gtrsim2$) and the column density is unconstrained.} with a broad Gaussian absorption feature. In the fitting procedure we restricted the energy of the line between $0.3$\,keV and $1.35$\,keV and its Gaussian $\sigma$ between $0$\,eV and $200$\,eV; the column density is varied between $\nh=0$\,cm$^{-2}$ and $5\times10^{21}$\,cm$^{-2}$, while the temperature of the blackbody components can assume values between $5$\,eV and $500$\,eV. For each observation, we fitted the model to the pn spectrum, to the combined MOS1 and MOS2 spectrum (labeled `MOS' in Table~\ref{tab_2bbg} and the two subsequent tables), and to the pn and MOS spectra simultaneously (labeled `EPIC' in Table~\ref{tab_2bbg} and the two subsequent tables). Finally, we performed multi-epoch simultaneous fits of all pn (5 spectra), MOS1/2 (10 spectra), and EPIC (6 spectra, comprising 5 pn and one combined MOS) exposures. 

For each fit in Table~\ref{tab_2bbg}, we list the reduced chi-square ($\chi^2_\nu$) and its null-hypothesis probability (NHP in \%), the column density $\nh$ in units of $10^{20}$\,cm$^{-2}$, the temperature of the cold $kT_1^\infty$ and hot $kT_2^\infty$ blackbody components in eV, the radiation radii $R_1^\infty$ and $R_2^\infty$ of each component (assuming a distance to the source of $d\equiv d_{300}=300$\,pc; \citealt[e.g.][]{2012PASA...29...98T}), the central energy of the absorption line (when constrained) or the corresponding $3\sigma$ upper limits, its Gaussian $\sigma$, and equivalent width $EW$ (all in eV), and the observed model flux in the energy band $0.2-12$\,keV, $f_{\rm X}$, in units of $10^{-12}$\,erg\,s$^{-1}$\,cm$^{-2}$. The model provides acceptable $\chi^2_\nu$ values in each epoch, with NHP between 14\% and 88\%. The somewhat large chi-square values and worse fit quality (NHP $<1\%$) of the multi-epoch fits could not be improved by the inclusion of an additional model component. 

In the left column of Figure~\ref{fig_2bbNSAgpar} we plot the results of Table~\ref{tab_2bbg} as a function of time, with $1\sigma$ errors. The best-fit parameters per instrument are consistent between pointings despite the systematic differences between the detectors. The column density is constrained and for the MOS and EPIC fits agrees within errors with the Galactic value in the direction of the source, $N_{\rm H}^{\rm gal}=(2.4-2.6)\times10^{20}$\,cm$^{-2}$ \citep[e.g.][]{2005A&A...440..775K,2013MNRAS.431..394W}. The pn camera measures twice as much absorption, as well as $7\%$ to $9\%$ softer temperatures with respect to MOS; the observed model flux is also $6\%$ lower. If $\nh$ is fixed to the Galactic value, the temperature of the two components and the observed model flux typically agree within $2.5\%$. 
Considering the simultaneous EPIC fit as an effective average between the detectors, the best-fit parameters are well constrained within ranges $\nh=(2.4-4)\times10^{20}$\,cm$^{-2}$, $kT_1^\infty=(64-68)$\,eV, $kT_2^\infty=(120-124)$\,eV, $R_1^\infty=(9.5-13)$\,km, $R_2^\infty=(1.0-1.2)$\,km, $\epsilon<(340-420)$\,eV, $\sigma=(100-120)$\,eV, $EW<(150-200)$\,eV, and $f_{X}=(6.4-6.9)\times10^{-12}$\,erg\,s$^{-1}$\,cm$^{-2}$. The model flux corrected for absorption (unabsorbed) is $F_{\rm X}=(1.2-1.7)\times10^{-11}$\,erg\,s$^{-1}$\,cm$^{-2}$.

Alternatively, the dataset can be as well fit (NHP $\sim10\%-80\%$) by a fully ionized neutron star hydrogen atmosphere model \citep[\textsf{\small nsa} in XSPEC;][]{1995ASIC..450...71P,1996A&A...315..141Z}, again modified by a broad Gaussian absorption line (Table~\ref{tab_nsag}). We tested non-magnetised ($B<10^8$\,G) and magnetised models with magnetic field values of $B=10^{12}$\,G and $B=10^{13}$\,G; in the fitting procedure, the neutron star mass and radius were at first fixed at the canonical values, $M=1.4$\,M$_\odot$ and $R=10$\,km, and then allowed to vary to check for an improved fit. In Table~\ref{tab_nsag} we list the results for a canonical neutron star with $B=10^{13}$\,G, which is the model that in most cases gave the highest NHP for each dataset. The unredshifted model effective temperature $T_{\rm eff}$ in K, the distance $d$ in pc, the parameters of the line ($\epsilon$, $\sigma$, $EW$), and the observed model flux $f_{\rm X}$, are also listed in Table~\ref{tab_nsag}.

\begin{figure}
\begin{center}
\includegraphics*[width=0.49\textwidth]{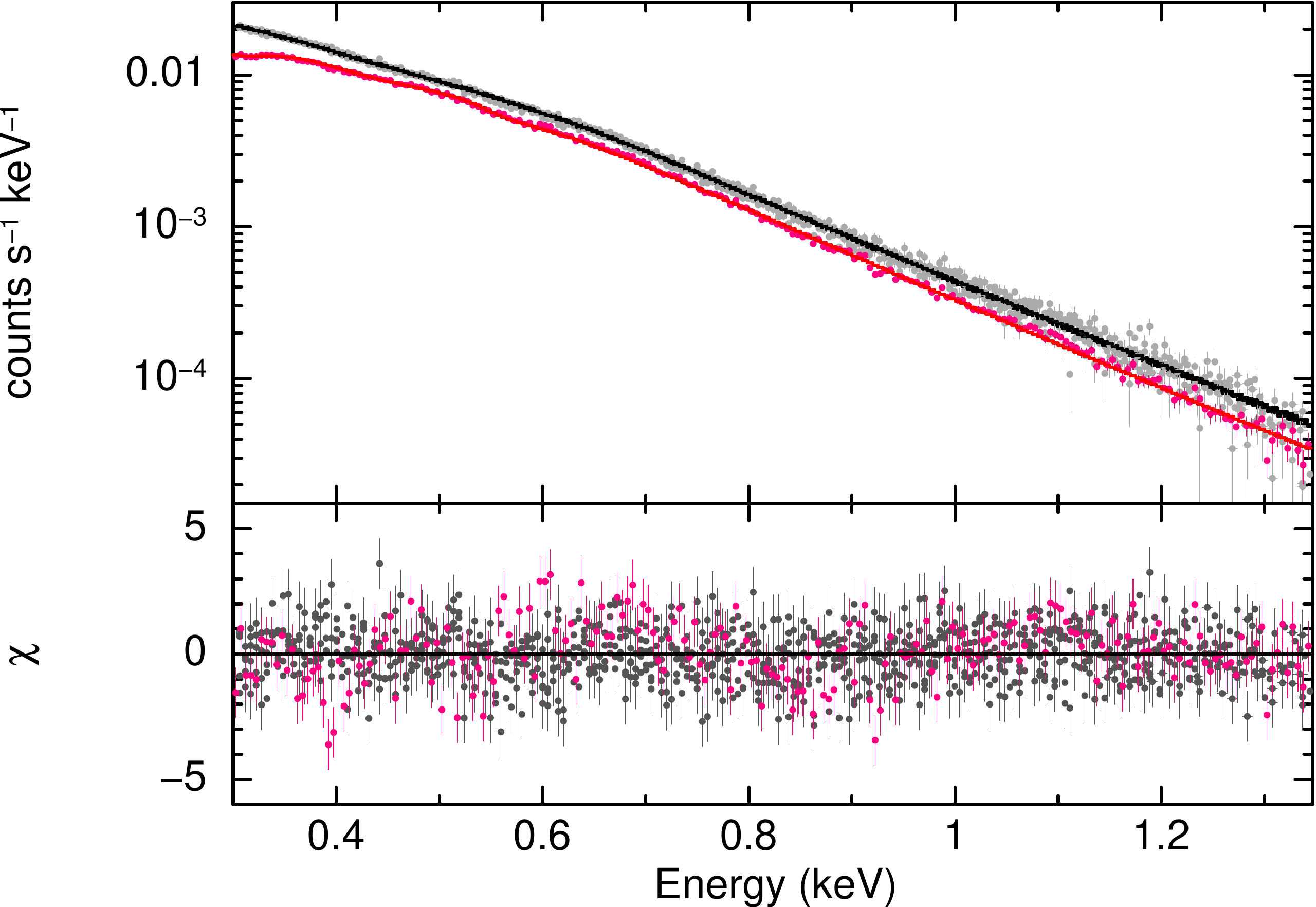}
\end{center}
\caption{Results of EPIC spectral fitting (see the text and Table~\ref{tab_nsag}, for details). We show the 5 pn and stacked MOS spectra (grey and magenta data points, respectively), fitted simultaneously by a fully ionized hydrogen neutron star atmosphere model with $B=10^{13}$\,G, $T_{\rm eff}=(5.16\pm0.17)\times10^5$\,K, and a broad Gaussian absorption line of $\sigma=91.4\pm0.4$\,eV at $\epsilon=373_{-7}^{+6}$\,eV (black and red solid lines).\label{fig_spec}}
\end{figure}

In the right column of Figure~\ref{fig_2bbNSAgpar} we plot the best-fit \textsf{\small nsa} parameters as a function of MJD (assuming for all epochs the results of the $B=10^{13}\,G$ fits with $M=1.4$\,M$_\odot$ and $R=10$\,km).
In comparison with the double blackbody model, the systematic differences between the detectors persist, however the measurements differ by a much smaller percentage -- $13\%$ in $\nh$, $\lesssim2\%$ in $T_{\rm eff}$, and $3\%$ in $f_{\rm X}$ -- and the overall consistency between epochs and instruments is improved (c.f.~overlapping shaded areas). The central energy of the line is well constrained, and also narrower than in the double blackbody case. 
Again considering the best-fit results of the EPIC fits we have: $\nh=(2-3)\times10^{20}$\,cm$^{-2}$, $T_{\rm eff}=(5.7-5.8)\times10^5$\,K, $B=10^{13}$\,G, $d=(124-129)$\,pc, $\epsilon=(360-400)$\,eV, $\sigma=(85-100)$\,eV, $EW=(55-60)$\,eV, and $f_{X}=(6.5-7.0)\times10^{-12}$\,erg\,s$^{-1}$\,cm$^{-2}$. The unabsorbed flux of this model is measured in a similar range as that of the double blackbody model, $F_{\rm X}=(1.2-1.4)\times10^{-11}$\,erg\,s$^{-1}$\,cm$^{-2}$. In Figure~\ref{fig_spec} we show this best-fit model folded to the EPIC dataset with residuals.

The neutron star distance derived from the \textsf{\small nsa} fits, $d\sim110-130$\,pc, is rather small in comparison to the range expected for the source, $d\sim300-400$\,pc \citep{2007Ap&SS.308..171P,2012PASA...29...98T}. While the fit is insensitive to the mass of the neutron star, unrealistically large neutron star radii, $R>20$\,km, and an overall poor fit quality, $\chi^2_\nu\sim1.7$, are obtained when the distance to the source is fixed at around the estimated value (see discussion in Section~\ref{sec_discussion}).

We explored other neutron star models where the atmosphere can be partially ionized and the size of the emission radius, $R_{\rm em}$, can be parametrised with respect to the neutron star physical radius \citep[\textsf{\small nsmaxg} in XSPEC;][]{2008ApJS..178..102H}.
For each epoch and instrument, as well as for the multi-epoch fits as before, we tested 19 absorbed \textsf{\small nsmaxg} models with $B=(0.01-30)\times10^{12}$\,G and $M=1.4$\,M$_\odot$. We first set the size of the emitting region to be the same as the neutron star radius; then we allowed this parameter to vary to smaller values to check for improved fits. As for the \textsf{\small nsa} models, we show in Table~\ref{tab_nsmaxg} the results with $B=10^{13}$\,G, which are the ones with the generally highest NHP and the most consistent parameters between epochs and instruments.  

The best-fit models (with somewhat comparable NHP as in the \textsf{\small 2bb} and \textsf{\small nsa} models, i.e., between $2\%$ and $89$\%) are for a neutron star atmosphere composed of hydrogen at a distance of less than $110$\,pc ($3\sigma$). All models consisting of mid-$Z$ element plasma (C, O, Ne) provided poor fit results. While the properties of the absorption line were found to be nearly identical to those of the fully ionized case, the temperature of the atmosphere is higher, and the radiation is roughly twice as much absorbed and inconsistent with the Galactic value. The size of the emission region was found to be slightly smaller than the canonical 10\,km of the \textsf{\small nsa} models, with $R_{\rm em}\sim8-9$\,km. The best-fit results of the EPIC fits are within: $\nh=(3-4)\times10^{20}$\,cm$^{-2}$, $T_{\rm eff}=(6.0-6.3)\times10^5$\,K, $B=10^{13}$\,G, $d<110$\,pc, $\epsilon=(340-400)$\,eV, $\sigma=(85-100)$\,eV, $EW=(50-65)$\,eV, and $f_{X}=(6.4-6.8)\times10^{-12}$\,erg\,s$^{-1}$\,cm$^{-2}$. The unabsorbed flux, $F_{\rm X}=(1.3-1.5)\times10^{-11}$\,erg\,s$^{-1}$\,cm$^{-2}$, is consistent with those of the other two previously discussed models. 

To break some of the degeneracy between the parameters and look for more physical results, we restricted the distance to the source within $d=100-600$\,pc, capped the column density at the Galactic value, and let the neutron star mass and radius vary within $M=0.5-2.5$\,M$_{\odot}$ and $R=5-15$\,km. However, the exercise led to generally worse fits and significant discrepancies between the best-fit parameters of pn and MOS. No other neutron star atmosphere model in XSPEC, nor the inclusion of a second (colder) component, provided acceptable fits.
\subsubsection{RGS data\label{sec_rgsspec}}
For the spectral analysis of RGS data we included four \xmm\ observations of the source performed in 2002/2003 in addition to the five 2012/AO14 observations, thus considerably extending the time span of the analysis in relation to that covered by the EPIC data (Tables~\ref{tab_dataexposure} and \ref{tab_rgsdata} in Section~\ref{sec_data}). The total analysed RGS dataset, of which the AO14 observations account for nearly 70\% in net exposure, amount to 18 RGS1/2 spectra and GTI-filtered exposures of 465\,ks and 457\,ks per detector. 

We used the EPIC source coordinates in each observation to generate the instrument spatial masks and energy filters with \textsf{\small rgsproc}. The GTI-filtered event lists were used to extract the source and background spectra in wavelength space using the tasks \textsf{\small rgsregions} and \textsf{\small rgsspectrum}, while response matrix files were produced with the SAS task \textsf{\small rmfgen}. Only the first-order spectra were analysed. To increase the signal-to-noise ratio each spectrum was rebinned into 0.165\,\AA\ wavelength channels. 
The defective channels of the RGS cameras\footnote{\xmm\ Calibration Technical Note 0030, issue 7.7; hereafter Gonz\`alez-Riestra et al.~(2018).}, which cover in first order the wavelength ranges of $11$\,\AA\ to $14$\,\AA\ in RGS1 and $20$\,\AA\ to $24$\,\AA\ in RGS2, were excluded from the spectral fitting. The total dataset amount to $3.992(20)\times10^4$ and $3.724(19)\times10^4$ counts ($15-30$\,\AA), respectively in each RGS1/2 camera, of which around $40\%$ can be ascribed to the background.

\begin{figure}
\begin{center}
\includegraphics*[width=0.49\textwidth]{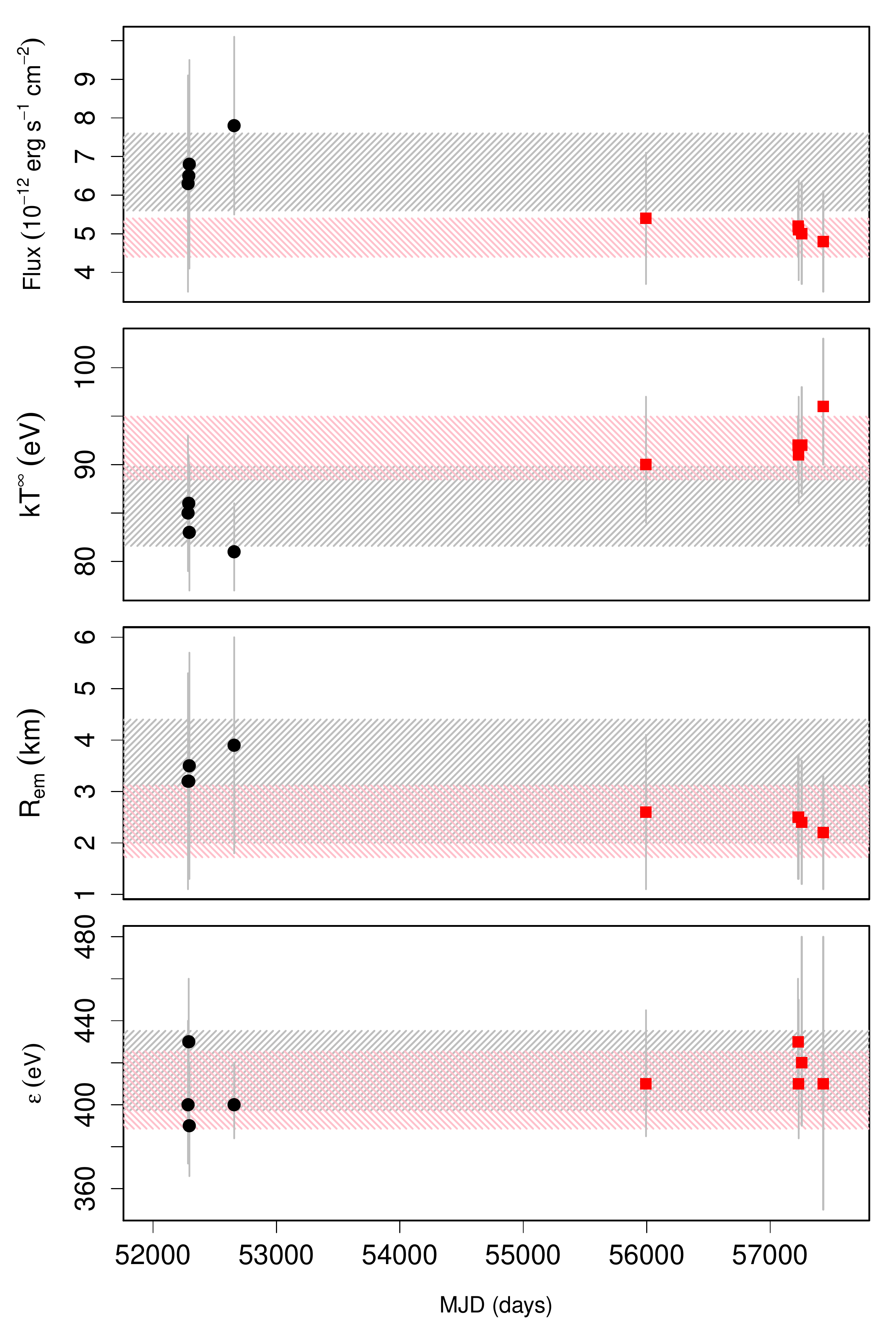}
\end{center}
\caption{Spectral parameters of \jsix\ as a function of time as measured in the RGS observations (data points; see the text and Table~\ref{tab_rgsvar} for details). The model fit to the observations in XSPEC is \textsf{\small tbabs(bbody-gauss)}. Circle (black) and square (red) symbols show the subgroups of `old' and `new' observations of the source, obtained respectively in 2002/2003 and after 2012. Note that the analysis of EPIC data (e.g.~Fig.~\ref{fig_2bbNSAgpar}) concern the observations in the `new' subgroup (MJD $>56,000$). Shaded areas are the results of the fits of RGS1/2 stacked spectra in the two subgroups, with $1\sigma$ standard deviations. \label{fig_rgspar}}
\end{figure}
We fitted each observation in XSPEC assuming a model (hereafter, the \textsf{\small bbgauss} model) consisting of an absorbed blackbody, modified by a Gaussian absorption feature with $\sigma=100$\,eV as found from the analysis of EPIC data; the column density was fixed to the Galactic value to better constrain the other model parameters. The RGS1/2 spectra were fitted simultaneously adopting a constant factor between the instruments. We note that the best-fit parameters from the fit of individual RGS1/2 spectrum agree well with each other in a given epoch. 

\begin{figure}
\begin{center}
\includegraphics*[width=0.49\textwidth]{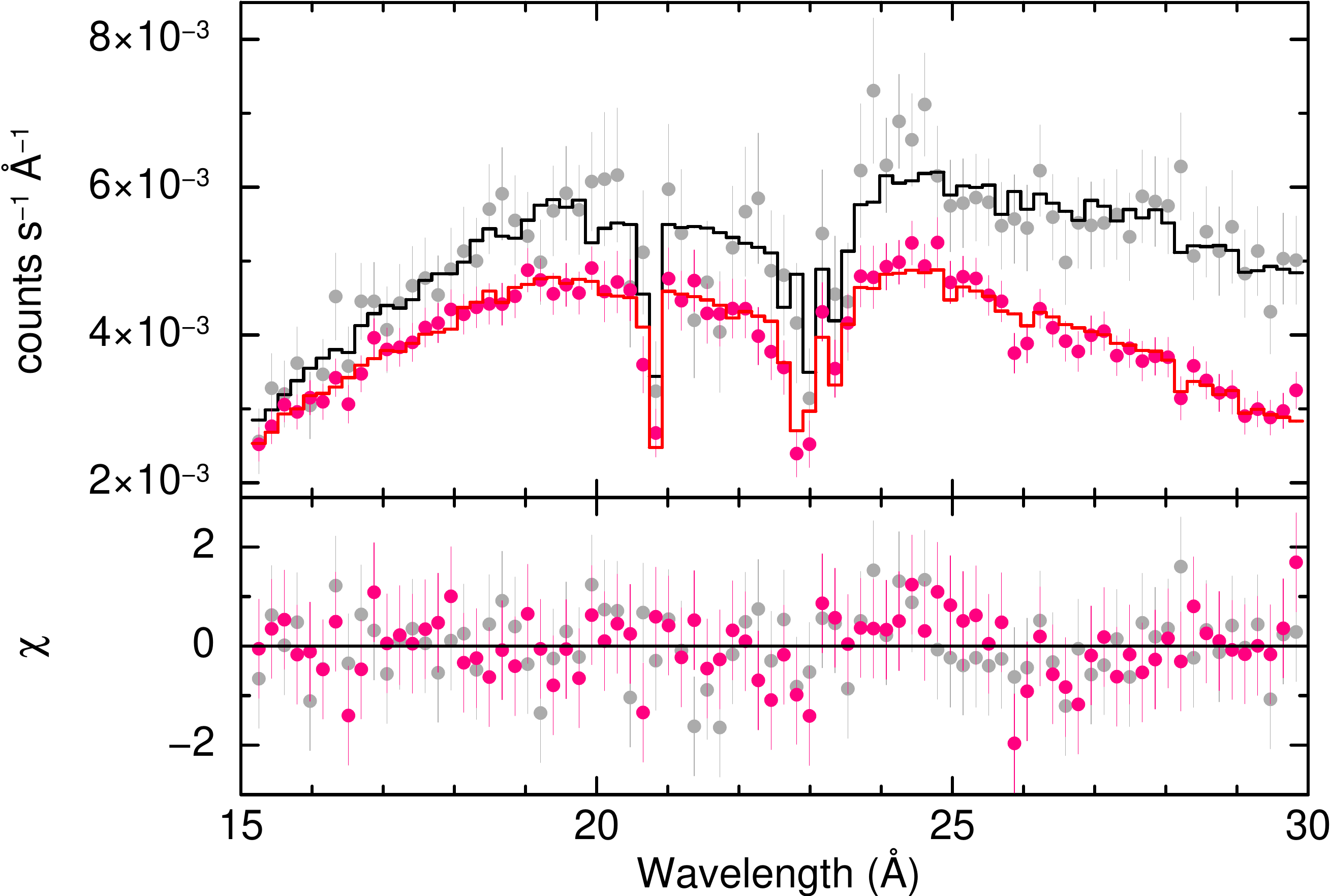}
\end{center}
\caption{Results of RGS spectral fitting (see the text and Table~\ref{tab_rgsvar}, for details). We show the stacked `old' and `new' RGS spectra (grey and magenta data points, respectively; the RGS1 and RGS2 spectra in each subgroup were co-added for plotting purposes). The model (solid black and red lines) fitted to the data in XSPEC is \textsf{tbabs(bbody-gauss)}. The best-fit parameters differ in the two subgroups (see text). The two absorption lines at around $\lambda=20-23$\,\AA\ are instrumental.\label{fig_specrgs}}
\end{figure}
The best-fit results of the \textsf{\small bbgauss} model are in Table~\ref{tab_rgsvar}. For each observation, we list the blackbody temperature $kT_\infty$ in eV, the radiation radius $R_{\infty}$ in km (assuming a source at $d_{300}$), the central energy $\epsilon$ and equivalent width $EW$ of the Gaussian absorption feature, and the unabsorbed source flux of the model $F_{\rm X}$, in the $0.2-12$\,keV energy band. 
The results suggest a possible trend of the parameters of the source: in comparison with the first four observations obtained in 2002 and 2003 and labeled (A-D) in Table~\ref{tab_rgsvar}, the more recent observations show a slight increase in temperature and a more pronounced decrease in the model normalization and flux (formally inconsistent with a constant term), at constant properties of the Gaussian absorption feature (Figure~\ref{fig_rgspar}). By contrast, the parameters of the source within these two subgroups are constant at the $2\%-3\%$ level in $kT_\infty$, $7\%-10\%$ in $R_\infty$, and $5\%-10\%$ in $F_{\rm X}$. The trends are seen in the spectral parameters of both RGS1/2 instruments. 

\begin{table}[t]
\caption{Results of the RGS spectral analysis
\label{tab_rgsvar}}
\centering
\begin{tabular}{@{}c r r r r r@{}} 
\hline\hline
\textsf{\small obsid}\tablefootmark{a} & \multicolumn{1}{c}{$kT_\infty$} & \multicolumn{1}{c}{$R_\infty$\tablefootmark{b}} & \multicolumn{1}{c}{$\epsilon$} & \multicolumn{1}{c}{$EW$} & \multicolumn{1}{c}{$F_{\rm X}$\tablefootmark{c}}\\
\cline{4-5} 
 & \multicolumn{1}{c}{(eV)} & \multicolumn{1}{c}{(km)} & \multicolumn{2}{c}{(eV)} & \multicolumn{1}{c}{} \\
 \hline
(A)  & $86(5)$        & $3.2_{-1.5}^{+1.8}$ & $430_{-24}^{+30}$ & $60^{+20}_{-16}$ & $6.5_{-1.4}^{+2.2}$\\
(B)  & $85_{-6}^{+8}$ & $3.2_{-1.9}^{+2.1}$ & $400_{-28}^{+40}$ & $45^{+30}_{-26}$ & $6.3_{-2.2}^{+2.8}$  \\
(C)  & $83_{-6}^{+7}$ & $3.5_{-1.9}^{+2.2}$ & $390_{-24}^{+30}$ & $55^{+24}_{-28}$ & $6.8_{-2.1}^{+2.7}$  \\
(D)  & $81_{-4}^{+5}$ & $3.9_{-1.9}^{+2.1}$ & $400_{-16}^{+20}$ & $85^{+5}_{-19}$ & $7.8_{-1.9}^{+2.3}$  \\
2012 & $90_{-6}^{+7}$ & $2.6_{-1.3}^{+1.5}$ & $410_{-25}^{+35}$ & $45^{+29}_{-25}$ & $5.4_{-1.3}^{+1.7}$\\
201  & $92(3)$        & $2.5_{-0.9}^{+1.0}$ & $430_{-23}^{+30}$ & $45^{+6}_{-26}$ & $5.2_{-0.7}^{+0.8}$\\
301  & $91_{-5}^{+6}$ & $2.5_{-1.1}^{+1.2}$ & $410_{-26}^{+40}$ & $55^{+6}_{-40}$ & $5.1_{-1.0}^{+1.3}$\\
401  & $92_{-5}^{+6}$ & $2.4_{-1.1}^{+1.2}$ & $420_{-30}^{+60}$ & $19^{+35}_{-15}$ & $5.0_{-1.0}^{+1.3}$\\
501  & $96_{-6}^{+7}$ & $2.2_{-1.0}^{+1.1}$ & $410_{-60}^{+70}$ & $<60$ & $4.8_{-1.0}^{+1.3}$\\
\hline
old  & $85.7_{-2.8}^{+3}$ & $3.20_{-1.2}^{+1.3}$ & $416_{-12}^{+15}$ & $50^{+2}_{-17}$ & $6.6_{-0.9}^{+1.0}$ \\
new  & $91.7_{-2.2}^{+2.4}$ & $2.42_{-0.7}^{+0.8}$ & $407_{-12}^{+14}$ & $60^{+12}_{-7}$ & $4.9(5)$ \\
\hline
\end{tabular}
\tablefoot{Errors are $1\sigma$ confidence levels. The model fitted to the data in XSPEC is \textsf{\small tbabs(bbody-gauss)}. The RGS1/2 spectra are fitted simultaneously in each epoch. The FWHM of the feature, $\sigma=100$\,eV, and the column density, $\nh=2.4\times10^{20}$\,cm$^{-2}$, are held fixed during spectral fitting. The chi-squared values are within $\chi^2\sim50-80$, for 135 degrees of freedom (NHP $\sim100$\,\%). 
\tablefoottext{a}{The observations are labeled as in Tables~\ref{tab_dataexposure} and \ref{tab_rgsdata}.}
\tablefoottext{b}{The radiation radius at infinity is computed for a source distance of $d\equiv d_{300}=300$\,pc.}
\tablefoottext{c}{The unabsorbed model flux is in units of $10^{-12}$\,erg\,s$^{-1}$\,cm$^{-2}$ in energy band $0.2-12$\,keV.} 
}
\end{table}
In \citet{2014A&A...563A..50P}, we investigated the constancy of the INS emission on the EPIC data performed between 2002 and 2012. Unfortunately, the analysis does not allow us to draw definite conclusions: while the MOS instruments are unsuited for long-term studies\footnote{XMM-Newton Calibration Technical Note 0018.}, the pn camera provides only one data point prior to 2012 for comparison, due to the heterogeneous dataset and background flares (Sections~\ref{sec_intro} and \ref{sec_data}). Nonetheless, an increase of blackbody temperature, consistent with what is observed in the RGS data, was then reported.  

To investigate the possibility of a long-term evolution on the parameters of \jsix, we co-added the spectra of the two subgroups in each RGS camera, using the SAS task \textsf{\small rgscombine}, and binned the results to 0.165\,\AA\ as before. The resulting grouped spectra (labeled `old' and `new' in Table~\ref{tab_rgsvar}) were then fitted simultaneously in XSPEC with the same \textsf{\small bbgauss} model, taking into account the co-added background and response files in each detector as usual. In Figure~\ref{fig_specrgs} we plot the `old' and `new' spectra  (grey and magenta data points) of \jsix, with the folded \textsf{\small bbgauss} model and fit residuals. The best-fit results as a function of time are plotted as shaded grey and pink areas in Figure~\ref{fig_rgspar}.

The results of this approach confirm the observed trend. With respect to the early pointings, we measure a $7\%$ higher temperature and a $25\%$ lower flux and smaller radiation radius in the observations performed after 2012, formally significant beyond the spectral errors. Nonetheless the significance of the variations is low: $1\sigma$ in $kT$ and $F_{X}$, while the other parameters are consistent within the rather large spectral errors (e.g.~the emission radius in Figure~\ref{fig_rgspar}). 

The RGS instruments suffer from a decline in sensitivity at long wavelengths, likely due to a build-up of of hydrocarbon contamination on the detector (\citealt{2015A&A...573A.128D}, Gonz\`alez-Riestra et al.~2018). 
Empirical corrections were first introduced in 2006 to take this and other effects into account in the calibrated model of the RGS1/2 effective areas, which are estimated to have an absolute accuracy of 10\%. 
Indeed, we observe a 18\% decrease in sensitivity in the effective area at long wavelengths between the old and new datasets (apparent in Figure~\ref{fig_specrgs}). 
Altogether, uncertainties from both the spectral model and other calibration issues, possibly not accounted for in the modeling of the effective area with time, may be responsible for the discrepancies on the parameters of \jsix\ reported here. 

The inclusion of a cold blackbody component in the \textsf{\small bbgauss} model, unlike for the EPIC data, is not satisfactory due to the large normalization required to fit the RGS spectra. If this is kept within reasonable limits (that is, corresponding to a $<10^{33}$\,erg\,s$^{-1}$ blackbody at $d=0.1-1$\,kpc), the quality of the fit is worsened; moreover, while there are no significant changes on the temperature of the hot component and on the parameters of the absorption line, the best-fit temperature of the cold component is very soft, $<30$\,eV, and the column density is 2 to 3 times higher than the Galactic value.

The evidence for a narrow absorption feature at energy $\epsilon\sim0.57$\,keV ($\lambda=21.5$\,\AA) in the RGS spectra of \jsix\ was first reported by \citet{2004ApJ...608..432V}. Similarly narrow features at around this wavelength have been identified in the RGS spectrum of the \msev\ INS \magzs\ \citep{2009A&A...497L...9H} and other thermally emitting INSs \citep{2012MNRAS.419.1525H}. To investigate the presence of the narrow feature in our dataset, we fitted each of the nine RGS1\footnote{The RGS2 data cannot be used due to the defective channels of the camera around the wavelength range of interest.} spectra individually, using an absorbed blackbody model and two Gaussian absorption lines. As the individual datasets do not have very high signal-to-noise, we fixed the column density to the Galactic value and set the energy and FWHM of the broad absorption line to the best parameters found consistently in the analysis of EPIC and RGS data ($\epsilon=410$\,eV and $\sigma=100$\,eV). Absorption features were then searched between 21\,\AA\ and 22.5\,\AA\ ($550-590$\,eV). 
\begin{table}[t]
\caption{Investigation of a narrow absorption feature in RGS1 data
\label{tab_rgsnarrow}}
\centering
\begin{tabular}{@{}c c r c r r c@{}} 
\hline\hline
\textsf{\small obsid}\tablefootmark{a} & \multicolumn{1}{c}{$S/N$} & \multicolumn{1}{c}{$kT_\infty$} & \multicolumn{1}{c}{$EW_1$} & \multicolumn{1}{c}{$\epsilon_2$} & \multicolumn{1}{c}{$\sigma_2$} & \multicolumn{1}{c}{$EW_2$}\\ 
\cline{3-7}
 & & \multicolumn{5}{c}{(eV)} \\
 \hline
(A)  & 45 & $84_{-4}^{+6}$ & $60_{-28}^{+22}$ & $571.9_{-1.7}^{+1.4}$ & $<5$ & $8$\\
(B)  & 40 & $83_{-4}^{+6}$ & $70$ & $-$ & $-$ & $<3$\\
(C)  & 40 & $83_{-4}^{+6}$ & $70$ & $-$ & $-$ & $<1.2$\\
(D)  & 50 & $84_{-4}^{+5}$ & $<100$ & $-$ & $-$ & $<2.1$\\
2012 & 50 & $91.1_{-6}^{+2.7}$ & $50_{-40}^{+20}$ & $552.0_{-1.4}^{+2.1}$ & $<10$ & $2.6$\\
201  & 90 & $94_{-4}^{+7}$ & $<60$ & $-$ & $-$ & $<9$\\
301  & 65 & $86_{-3}^{+5}$ & $50$ & $-$ & $-$ & $<4$\\
401  & 60 & $92_{-5}^{+8}$ & $<50$ & $<560$ & $<10$ & $2.4$\\
501  & 60 & $96_{-10}^{+5}$ & $30$ & $-$ & $-$ & $<9$\\
\hline
old  & 90 & $83.9_{-2.3}^{+2.7}$ & $70$ & $575(4)$ & $5.1_{-2.2}^{+4}$ & $3$\\
new  & 150 & $93(3)$ & $35$ & $<560$ & $16_{-10}^{+12}$ & $4$\\
\hline
\end{tabular}
\tablefoot{Errors are $1\sigma$ confidence levels. The model fitted to the data in XSPEC is \textsf{\small tbabs(bbody-gauss-gauss)}. The subscripts `1' and `2' in the Table labels refer to the properties of the broad and narrow features, respectively. The energy, $\epsilon_1=410$\,eV, and FWHM, $\sigma_1=100$\,eV, of the broad absorption feature, as well as the hydrogen column density, $\nh=2.4\times10^{20}$\,cm$^{-2}$, are held fixed during spectral fitting. The chi-squared values are within $\chi^2\sim25-50$, for 76 degrees of freedom (NHP $\sim100$\,\%).
\tablefoottext{a}{The observations are labeled as in Tables~\ref{tab_dataexposure} and \ref{tab_rgsdata}.}
}
\end{table}

The results are summarised in Table~\ref{tab_rgsnarrow}. For each fit, we show the signal-to-noise $S/N$ of each spectrum, the blackbody temperature $kT$ of the source and the equivalent width of the broad absorption feature $EW_1$; the energy $\epsilon_2$, FWHM $\sigma_2$, and equivalent width $EW_2$ of the narrow feature -- when constrained, or their corresponding $1\sigma$ upper limits -- are also listed. 

The best-fit spectral parameters are consistent with those of Table~\ref{tab_rgsvar}, showing that the inclusion of the narrow feature is not statistically required in most cases.
Considering the number of trials ($40$) in the searched wavelength range, the evidence for the narrow feature is only significant in observation (A). Remarkably, there is no evidence for a narrow feature within $21-22.5$\,\AA\ in the longest observation of the source (labeled `201'), which has a much higher $S/N$ than the others. The same analysis carried out in the co-added `old' and `new' spectra confirm that additional features are absent in the most recent pointings. 
\section{Discussion\label{sec_discussion}}
The \msev\ have been considered a rather homogeneous group of cooling neutron stars, displaying similar ages, temperatures, and timing properties. The source \magos\ stood out in that it could be slowing down at a fast rate, indicating a high dipolar field -- the highest amongst the group -- and a possible evolution from a magnetar \citep{2014A&A...563A..50P}. The analysis of our dedicated \xmm\ large programme does not confirm the previous results (Section~\ref{sec_timing}). Due to the energy-dependent nature of the previously detected modulation, we performed extensive high-resolution periodicity searches allowing for moderate changes of pulsed fraction and the optimal energy range and signal-to-noise ratio for detection, for a reasonably wide range of spin down values. No significant signal resulted from the analysis: unless considerable changes of pulsed fraction have taken place since 2012, the deepest upper limit of $1.33(6)\%$ ($4\sigma$), in the relevant frequency range, conservatively rules out the $3.39$\,s modulation. Moreover, in the full frequency range allowed by the timing resolution of the EPIC cameras, blind searches revealed no other periodic signals with $p_{\rm f}\gtrsim1.5\%$ ($3\sigma$; $0.3-1.35$\,keV), thus considerably improving previous estimates for pulsations with $P>0.15$\,s. Similarly low $3\sigma$ upper limits, within $1.8\%$ and $4\%$, are obtained in the same period range in narrow (100\,eV to 600\,eV wide) energy intervals, defined according to the source's signal-to-noise ratio. 

With over $10^6$ EPIC counts, the unprecedented photon statistics of the new dataset allowed the deepest to-date investigation of the X-ray emission of the source (Section~\ref{sec_epicspec}). We found that, altogether, no theoretical model available in XSPEC can provide a fully satisfactory physical description of the spectrum of the neutron star. While statistically acceptable fits are obtained for the individual epochs, multi-epoch fits including the spectra of all EPIC cameras have null-hypothesis probabilities of less than $1\%$, which may at least partially be ascribed to cross-calibration uncertainties. Best results were obtained when fitting the data with either a double-blackbody (\textsf{\small 2bb}) or a magnetised neutron star atmosphere model consisting of hydrogen (\textsf{\small nsa} and \textsf{\small nsmaxg}, with $B=10^{13}$\,G and $M=1.4$\,M$_\odot$), in either case modified by a broad Gaussian absorption feature as previously reported in the literature. 

No significant evidence of spectral variability is measured in the 2012--2016 time frame covered by the analysis of EPIC data. The overall consistency of the parameters (between epochs and EPIC instruments) was optimal for the atmosphere models, with systematic errors of $13\%$ in column density, $2\%$ in temperature, and $3\%$ in flux. In particular, a canonical \textsf{\small nsa} model with $B=10^{13}$\,G much better constrains the column density toward the source and the properties of the absorption line in relation to the \textsf{\small 2bb} model. Best-fit distances around $d\sim130$\,pc are, on the other hand, inconsistently smaller than that estimated for the source, $300\pm50$\,pc. The model consisting of a partially ionized hydrogen neutron star atmosphere (\textsf{\small nsmaxg}) provides even smaller distances ($3\sigma$ upper limits below $110$\,pc) for a $R_{\rm em}\sim(8-9)$\,km emitting region on the neutron star, while the derived column density is nearly two times the Galactic value in the direction of the source. 

Considering the \textsf{\small nsa} model of the multi-epoch pn fits (which is the result with the highest NHP among the multi-epoch fits), the spectral parameters of the source are constrained as $\nh=2.45(14)\times10^{20}$\,cm$^{-2}$, $T_{\rm eff}=5.81(3)\times10^{5}$\,K, $\epsilon=385\pm10$\,eV, $\sigma=86.8\pm0.3$\,eV, $EW=55\pm3$\,eV, and $f_{\rm X}=6.842(9)\times10^{-12}$\,erg\,s$^{-1}$\,cm$^{-2}$ ($0.2-12$\,keV). In contrast to previous analysis, we did not find that the inclusion of other model components (in particular, additional lines in absorption) were statistically justified or could significantly improve the results of the multi-epoch fits. Other up-to-date, fully and partially ionized neutron star atmosphere models, consisting of different elemental compositions and with non-canonical neutron star mass and radius, did not provide better fits than the models described above.

The typical distance derived from the best-fit atmosphere models, around 130\,pc, is smaller than the range expected for the source, $d=350\pm50$\,pc \citep[Section~\ref{sec_epicspec}]{2007Ap&SS.308..171P}. This range is derived from the fitted hydrogen column density, assuming a blackbody model with three Gaussian lines in absorption, and a three-dimensional description of the distribution of the interstellar medium in the direction of the source (which, according to the authors, should be reliable up to $\sim270$\,pc). Based on kinematic arguments, \citet{2012PASA...29...98T} applied this expected range and the observed proper motion of the source \citep{2005A&A...429..257M,2006A&A...457..619Z} to trace back the neutron star trajectory and determine its likely birthplace, using possible associations with runaway massive stars and the observed abundance of heavy elements as further evidence to narrow down the most likely solutions. These predict a current distance of $d=300-370$\,pc if the neutron star was born less than 0.5\,Myr ago in a nearby supernova explosion.
\subsection{The viewing geometry and presence of hotspots\label{sec_res}}
Magnetic fields in the range observed in the \msev\ are expected to produce large temperature variations on the neutron star surface, due to the anisotropic electron conductivity and heat transport in the stellar envelope and crust \citep[e.g.][]{2004A&A...426..267G,2006A&A...451.1009P,2013MNRAS.434.2362P}. In this case, strong pulsed flux variations are expected at the neutron star spin period unless the source is observed from a particularly unfavourable geometry: either if the angle between the neutron star spin axis and the line-of-sight $i$ is sufficiently small, or if the regions of higher temperature (hotspots) are located at a very small angle $\theta_B$ in relation to the neutron star rotation axis (that is, the magnetic and spin axes of the star are nearly co-aligned). Therefore, the stringent pulsed fraction limits from the timing analysis can be used to verify the viability of the \textsf{\small 2bb} model and probe the presence of hotspots on the surface of \jsix\ \citep[e.g., see the case of the thermally emitting central neutron star in the supernova remnant \hess1731,][]{2017A&A...600A..43S}.
\begin{figure}
\begin{center}
\includegraphics*[width=0.495\textwidth]{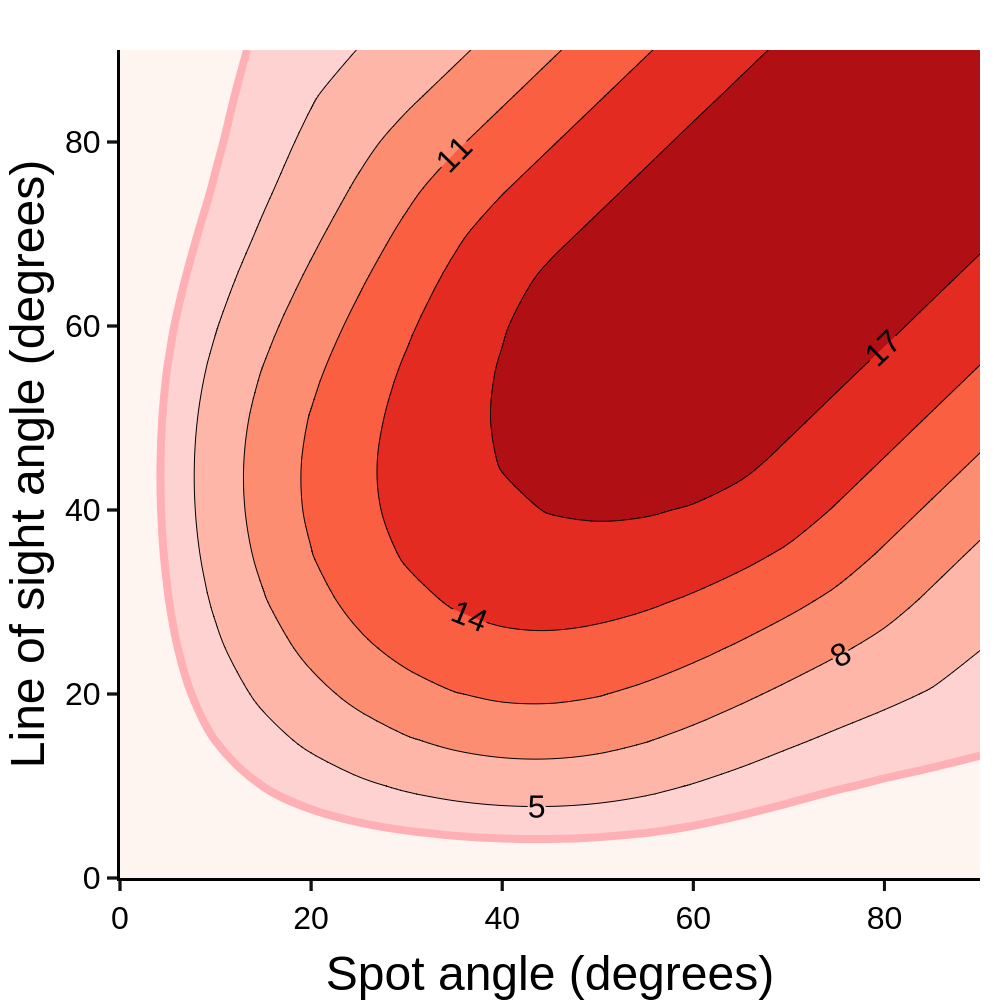}
\end{center}
\caption{Contours of constant pulsed fraction assuming the best-fit parameters of the pn double blackbody model ($kT_{\rm ns}=60.9_{-1.5}^{+1.7}$\,eV, $R_{\rm ns}^\infty=16.4_{-2.2}^{+2.9}$\,km, $kT_{\rm sp}=117.0(8)$\,eV, $R_{\rm sp}^\infty=1.35_{-0.18}^{+0.24}$\,km). Count rates are computed in the $0.5-1.35$\,keV energy band to exclude the effects of the Gaussian absorption line and pulsed fraction labels are given in \%. Darker colours denote higher pulsed fractions. The allowed parameter space of the viewing geometry constrained by the timing analysis lies below the thick pink line ($p_{\rm f}\le2.78(16)\%$; Table~\ref{tab_upperlim}). \label{fig_contour}}
\end{figure}

With this goal we considered the model of a slowly rotating neutron star, observed at an inclination angle $i$, with two identical polar hotspots located at an angle $\theta_B$ with respect to the spin axis \citep[see, e.g.][for a full description of the model]{1995ApJ...442..273P,2005A&A...441..597S,2010A&A...522A.111S}. Light bending in the vicinity of the neutron star follows the relation between the local angle of the emitted photon and its escape direction and depends on the compactness of the neutron star (given by the ratio between the neutron star and the Schwarzschild radius $r_{\rm g}=R_{\rm ns}c^2(2GM_{\rm ns})^{-1}$; \citealt{2002ApJ...566L..85B}). 
The temperature of the neutron star surface, $kT_{\rm ns}^\infty=60.9_{-1.5}^{+1.7}$\,eV, and of the hotspots, $kT_{\rm sp}^{\infty}=117.0\pm0.8$\,eV, are assumed from the best-fit \textsf{\small 2bb} pn model, which is the one with the highest NHP (Table~\ref{tab_2bbg}). For a neutron star distance of $d=300^{+50}_{-40}$\,pc \citep{2012PASA...29...98T}, the model normalisations set the corresponding sizes of the emission regions as $R_{\rm ns}^\infty=16.4_{-2.2}^{+2.9}$\,km and $R_{\rm sp}^\infty=1.35_{-0.18}^{+0.24}$\,km. 
With these values the angular size of the spots and the compactness of the star are fixed as $\theta_{\rm sp}=4.7^\circ$ and $r_{g}=2.8$, respectively, assuming a 1.5\,M$_\odot$ neutron star. 

For a particular viewing geometry ($i$, $\theta_B$), the photon flux at a given rotation phase results from the sum of the visible individual area elements of the neutron star surface, assuming blackbody emission and taking into account the light bending. The flux is corrected for the interstellar absorption of a equivalent column density of $\nh^{\sf 2bb}=4.5\times10^{20}$\,cm$^{-2}$ and then folded with the EPIC pn response to give the source count rate at the $0.5-1.35$\,keV energy band. The energy band is chosen to minimise the effects of the broad absorption feature in the emitted spectrum. 

With this method we computed an extensive grid of lightcurves for $(\theta_B,i)$ within $(0^\circ,0^\circ)$ and $(90^\circ,90^\circ)$ and computed the pulsed fraction for each orientation as:

\begin{displaymath} 
p_{\rm f}=\frac{CR_{\rm max}-CR_{\rm min}}{CR_{\max}+CR_{\rm min}}
\end{displaymath}

In Figure~\ref{fig_contour} we plot the resulting $p_{\rm f}$ map in the $(\theta_B,i)$ plane. The maximum pulsed fraction obtained for the \textsf{\small 2bb} model is about 20\%. The region allowed by the limits of the timing analysis lies below the thick pink line corresponding to $p_{\rm f}=2.78(16)$\% (Table~\ref{tab_upperlim}; $0.5-1.35$\,keV). Integrating over all possible random orientations of line-of-sight inclination and spot angles we obtain a small likelihood ($\sim$1.9\%) that we do not see pulsations from the source due to the particularly unfavourable viewing geometry, if the \textsf{\small 2bb} model is correct.  
\subsection{The energy distribution\label{sec_sed}}
Thermal emission from INSs is expected to originate immediately at the surface, with the bulk of the energy flux peaking in the soft X-ray band. In principle, by confronting the observed spectra and light curves with theoretical models for neutron star thermal radiation, it should be possible to derive the surface temperature, magnetic field, gravitational acceleration and chemical composition: if distances are known, then the stellar mass, radius and the equation of state of neutron star interior could be constrained as well \citep[see][for recent reviews on neutron star atmosphere models and up-to-date astrophysical constraints on the equation of state of nuclear matter]{2015SSRv..191..171P,2016ARA&A..54..401O}. Since their discovery in the All-Sky Survey of the \ros\ satellite \citep{1999A&A...349..389V}, the \msev\ have been regarded as the closest-to-perfect candidates for testing neutron star emission models, due to a combination of bright thermal emission, proximity, independent distance estimates\footnote{From HST parallaxes (in two cases) and kinematic studies \citep[e.g.][and references therein]{2010ApJ...724..669W,2010MNRAS.402.2369T}.}, and a lack of significant magnetospheric or accretion activity. 

In practice, progress has been hampered by uncertainties on the chemical composition of the atmosphere and the phase state of the stellar surface, as well as by the lack of understanding on the magnetic field and temperature distributions \citep[e.g.][]{2006MNRAS.366..727Z,2007Ap&SS.308..191V,2010A&A...522A.111S}. The presence of lines adds to this complexity as, although believed to be related with the star's magnetic field, they have no unique physical interpretation. To explain the emitted radiation and equivalent widths of the lines in the phase-resolved spectrum of the \msev\ \magot, \citet{2010A&A...522A.111S} favoured a model where a partially ionized, optically thin atmosphere above the condensed surface must be present \citep[see also][]{2003A&A...408..323M,2007MNRAS.375..821H}. Using this model, \citet{2011A&A...534A..74H} derived the temperatures of the X-ray emitting areas and the magnetic field intensity at the poles; moreover, they could constrain the compactness of the neutron star and the gravitational redshift on the surface, suggesting a very stiff equation of state. Similar conclusions were reached for the \msev\ \magzs\ \citep{2017A&A...601A.108H}. These results are only marginally compatible with the most favoured range of the true radius of a 1.5\,M$_\odot$ neutron star, $10-11.5$\,km, from the analysis of \citet{2016ARA&A..54..401O}. New generation X-ray missions, in particular the Neutron Star Interior Composition Explorer Mission \citep[NICER,][]{2012SPIE.8443E..13G}, together with more accurate distances from the GAIA satellite \citep{2016A&A...595A...1G}, will certainly improve the constraints on neutron star mass and radius from astrophysical observations of, for example, millisecond pulsars in globular clusters.

Recently, \citet{2014MNRAS.443...31V} showed that in some cases (as for the \msev\ \magze; \citealt{2004A&A...424..635H}) the deviations found in the spectra of thermally emitting INSs may be induced simply by the inhomogeneous temperature distribution on the surface. While the effect is unlikely to account for all cases of sources with reported spectral features, the interesting result is that the anisotropic temperature distribution can give way to ``spurious'' spectral features to be claimed. We can safely exclude this possibility for the absorption feature in \jsix, which cannot be accommodated by a multi-temperature energy distribution. 

All \msev\ INSs have detected optical, ultraviolet, or infrared counterparts \citep[see][for references and limits]{2011ApJ...736..117K,2014ApJS..215....3P,2018ApJ...865....1P}. Interestingly, the extrapolation to longer wavelengths of the best-fit model inferred from X-rays -- including, for the case of \jsix, both the double temperature blackbody and atmosphere models discussed here -- falls below the actual detected fluxes: this is known as the `optical excess', and is observed in all \msev\ INSs. The optical excess of \jsix\ deviates significantly from the expected Rayleigh-Jeans slope of the spectra and can be described by a rather flat power-law \citep{2011ApJ...736..117K}.
The origin of the excess flux might rely on atmospheric effects, magnetospheric emission, or resonant scattering. In particular, if the X-ray and optical/UV emission came from different regions on the surface \citep[e.g.][]{2002ApJ...580.1043B}, we might expect correlations between the amount of optical excess and the X-ray pulsed fraction, which are not verified. The possible presence of fossil fallback disks, of a faint pulsar wind nebula, `bare' neutron star surfaces, and other alternative scenarios remain open \citep[see][for a discussion]{2004ApJ...603..265T,2017MNRAS.470.1253E,2017ApJ...837...81W,2018ApJ...865....1P}. 
\subsection{Narrow absorption feature\label{sec_RGSnarrow}}
In combination with the existing archival RGS data, the AO14 campaign accumulated over $450$\,ks of exposure time on \jsix, increasing the available data by $70\%$. The good statistics allowed a detailed analysis of the narrow feature at $\epsilon\sim570$\,eV, previously reported in the literature. The investigation shows that the feature is only significantly detected in one early epoch (the 2002 observation first analysed by \citealt{2004ApJ...608..432V}), or when this observation is combined with the archival data obtained prior to the AO14 campaign \citep{2012MNRAS.419.1525H,2014A&A...563A..50P}. The feature is definitely not present in the AO14 observations, while evidence of a less significant narrow feature is present at energy $\epsilon=550$\,eV in the 2012 observation. Consistently with these results, the analysis of the two grouped spectra of early and recent observations constrain the presence of the narrow feature only in the first subgroup (Section~\ref{sec_rgsspec}). 

\citet{2009A&A...497L...9H} discusses the possible physical interpretation of a similar feature detected in the co-added RGS spectrum of \magzs, which was later confirmed by the analysis of \chan\ LETG data \citep{2012MNRAS.419.1525H}. Their analysis favours a blend of highly ionized oxygen originating in the ambient medium of the INS, possibly a high density nearby cloud which could contribute to the source's optical excess. Nonetheless, a interstellar or atmosphere origins cannot be ruled out. At least for \jsix, the narrow and transient nature of the feature disfavours an atmospheric origin. 

Phase-dependent narrow absorption features have been reported in \xmm\ observations of the \msev\ INSs \magzs\ and \magot\ \citep{2015ApJ...807L..20B,2017MNRAS.468.2975B}, a work motivated by the detection of variable cyclotron lines detected in the spectra of two `low magnetic field' magnetars \citep{2013Natur.500..312T,2016MNRAS.456.4145R}. These results give support for the presence of strong, confined magnetic field components close to the stellar surface and a complex field topology. In contrast, the features in the spectra of the two \msev\ INSs are intrinsically different in that they do not vary in energy, are detected at much lower energy, and are also seem to be stable and lasting over long timescales. 
\section{Summary and conclusions\label{sec_summary}}
We report here the results of a \xmm\ large programme on the thermally emitting isolated neutron star \magos. The project aimed to gain a deeper understanding of the timing and spectral properties of the source, through a detailed analysis of its X-ray emission. The neutron star is of particular scientific interest as a source that could potentially bridge the evolutionary gap between the groups of nearby thermally emitting sources dubbed the `magnificent seven' and the young and energetic magnetars. Due to the lack of detected pulsations, our science goals could only be partially completed. Nonetheless, the deep upper limits derived from our analysis were used to put stringent constraints on the viewing geometry and the presence of hot spots on the surface. Detailed phase-averaged medium and high-resolution spectroscopy constrains atmosphere neutron star models and the properties of the cyclotron line in the spectrum of the neutron star with unprecedented statistics. The non-detection of the narrow absorption feature at $\epsilon=570$\,eV reported in previous epochs also disfavours an atmospheric origin. 
\begin{acknowledgements}
We thank the anonymous referee for useful comments and suggestions which helped to improve the paper.
We thank Norbert Schartel and the staff of the \xmm\ Science Operation Center, in particular Jan-Uwe Ness, for their great support in the scheduling of these time-constrained observations. We acknowledge the use of the ATNF Pulsar Catalogue (\texttt{http://www.atnf.csiro.au/people/pulsar/psrcat}).
This work was supported by the Deutsches Zentrum f\"ur Luft- und Raumfahrt (DLR) under grant 50 OR 1511. 
\end{acknowledgements}
\bibliographystyle{aa}
\bibliography{ao14ref}

\hyphenation{Post-Script Sprin-ger}
\begin{thebibliography}{83}
\expandafter\ifx\csname natexlab\endcsname\relax\def\natexlab#1{#1}\fi

\bibitem[{{Arnaud}(1996)}]{1996ASPC..101...17A}
{Arnaud}, K.~A. 1996, in Astronomical Society of the Pacific Conference Series,
  Vol. 101, Astronomical Data Analysis Software and Systems V, ed. G.~H.
  {Jacoby} \& J.~{Barnes}, 17

\bibitem[{{Beloborodov}(2002)}]{2002ApJ...566L..85B}
{Beloborodov}, A.~M. 2002, \apjl, 566, L85

\bibitem[{{Beloborodov} \& {Li}(2016)}]{2016ApJ...833..261B}
{Beloborodov}, A.~M. \& {Li}, X. 2016, \apj, 833, 261

\bibitem[{{Borghese} {et~al.}(2015){Borghese}, {Rea}, {Coti Zelati}, {Tiengo},
  \& {Turolla}}]{2015ApJ...807L..20B}
{Borghese}, A., {Rea}, N., {Coti Zelati}, F., {Tiengo}, A., \& {Turolla}, R.
  2015, \apjl, 807, L20

\bibitem[{{Borghese} {et~al.}(2017){Borghese}, {Rea}, {Coti Zelati}, {Tiengo},
  {Turolla}, \& {Zane}}]{2017MNRAS.468.2975B}
{Borghese}, A., {Rea}, N., {Coti Zelati}, F., {et~al.} 2017, \mnras, 468, 2975

\bibitem[{{Braje} \& {Romani}(2002)}]{2002ApJ...580.1043B}
{Braje}, T.~M. \& {Romani}, R.~W. 2002, \apj, 580, 1043

\bibitem[{{Buccheri} {et~al.}(1983){Buccheri}, {Bennett}, {Bignami}, {Bloemen},
  {Boriakoff}, {Caraveo}, {Hermsen}, {Kanbach}, {Manchester}, {Masnou},
  {Mayer-Hasselwander}, {{\"O}zel}, {Paul}, {Sacco}, {Scarsi}, \&
  {Strong}}]{1983A&A...128..245B}
{Buccheri}, R., {Bennett}, K., {Bignami}, G.~F., {et~al.} 1983, \aap, 128, 245

\bibitem[{{Coti Zelati} {et~al.}(2018){Coti Zelati}, {Rea}, {Pons}, {Campana},
  \& {Esposito}}]{2018MNRAS.474..961C}
{Coti Zelati}, F., {Rea}, N., {Pons}, J.~A., {Campana}, S., \& {Esposito}, P.
  2018, \mnras, 474, 961

\bibitem[{{de Vries} {et~al.}(2015){de Vries}, {den Herder}, {Gabriel},
  {Gonzalez-Riestra}, {Ibarra}, {Kaastra}, {Pollock}, {Raassen}, \&
  {Paerels}}]{2015A&A...573A.128D}
{de Vries}, C.~P., {den Herder}, J.~W., {Gabriel}, C., {et~al.} 2015, \aap,
  573, A128

\bibitem[{{den Herder} {et~al.}(2001){den Herder}, {Brinkman}, {Kahn},
  {Branduardi-Raymont}, {Thomsen}, {Aarts}, {Audard}, {Bixler}, {den Boggende},
  {Cottam}, {Decker}, {Dubbeldam}, {Erd}, {Goulooze}, {G{\"u}del}, {Guttridge},
  {Hailey}, {Janabi}, {Kaastra}, {de Korte}, {van Leeuwen}, {Mauche},
  {McCalden}, {Mewe}, {Naber}, {Paerels}, {Peterson}, {Rasmussen}, {Rees},
  {Sakelliou}, {Sako}, {Spodek}, {Stern}, {Tamura}, {Tandy}, {de Vries},
  {Welch}, \& {Zehnder}}]{2001A&A...365L...7D}
{den Herder}, J.~W., {Brinkman}, A.~C., {Kahn}, S.~M., {et~al.} 2001, \aap,
  365, L7

\bibitem[{{Ertan} {et~al.}(2017){Ertan}, {{\c{C}}al{\i}șkan}, \&
  {Alpar}}]{2017MNRAS.470.1253E}
{Ertan}, {\"U}., {{\c{C}}al{\i}șkan}, {\c{S}}., \& {Alpar}, M.~A. 2017,
  \mnras, 470, 1253

\bibitem[{{Esposito} {et~al.}(2018){Esposito}, {Rea}, \&
  {Israel}}]{2018arXiv180305716E}
{Esposito}, P., {Rea}, N., \& {Israel}, G.~L. 2018, arXiv e-prints
  [\eprint[arXiv]{1803.05716}]

\bibitem[{{Evans} {et~al.}(2010){Evans}, {Primini}, {Glotfelty}, {Anderson},
  {Bonaventura}, {Chen}, {Davis}, {Doe}, {Evans}, {Fabbiano}, {Galle}, {Gibbs},
  {Grier}, {Hain}, {Hall}, {Harbo}, {(Helen He}, {Houck}, {Karovska},
  {Kashyap}, {Lauer}, {McCollough}, {McDowell}, {Miller}, {Mitschang},
  {Morgan}, {Mossman}, {Nichols}, {Nowak}, {Plummer}, {Refsdal}, {Rots},
  {Siemiginowska}, {Sundheim}, {Tibbetts}, {Van Stone}, {Winkelman}, \&
  {Zografou}}]{2010ApJS..189...37E}
{Evans}, I.~N., {Primini}, F.~A., {Glotfelty}, K.~J., {et~al.} 2010, \apjs,
  189, 37

\bibitem[{{Gaia Collaboration} {et~al.}(2016){Gaia Collaboration}, {Prusti},
  {de Bruijne}, {Brown}, {Vallenari}, {Babusiaux}, {Bailer-Jones}, {Bastian},
  {Biermann}, {Evans}, \& et~al.}]{2016A&A...595A...1G}
{Gaia Collaboration}, {Prusti}, T., {de Bruijne}, J.~H.~J., {et~al.} 2016,
  \aap, 595, A1

\bibitem[{{Gendreau} {et~al.}(2012){Gendreau}, {Arzoumanian}, \&
  {Okajima}}]{2012SPIE.8443E..13G}
{Gendreau}, K.~C., {Arzoumanian}, Z., \& {Okajima}, T. 2012, in \procspie, Vol.
  8443, Space Telescopes and Instrumentation 2012: Ultraviolet to Gamma Ray,
  844313

\bibitem[{{Geppert} {et~al.}(2004){Geppert}, {K{\"u}ker}, \&
  {Page}}]{2004A&A...426..267G}
{Geppert}, U., {K{\"u}ker}, M., \& {Page}, D. 2004, \aap, 426, 267

\bibitem[{{Goldreich} \& {Reisenegger}(1992)}]{1992ApJ...395..250G}
{Goldreich}, P. \& {Reisenegger}, A. 1992, \apj, 395, 250

\bibitem[{{Groth}(1975)}]{1975ApJS...29..285G}
{Groth}, E.~J. 1975, \apjs, 29, 285

\bibitem[{{Haberl}(2007)}]{2007Ap&SS.308..181H}
{Haberl}, F. 2007, \apss, 308, 181

\bibitem[{{Haberl} {et~al.}(2004){Haberl}, {Motch}, {Zavlin}, {Reinsch},
  {G{\"a}nsicke}, {Cropper}, {Schwope}, {Turolla}, \&
  {Zane}}]{2004A&A...424..635H}
{Haberl}, F., {Motch}, C., {Zavlin}, V.~E., {et~al.} 2004, \aap, 424, 635

\bibitem[{{Hambaryan} {et~al.}(2009){Hambaryan}, {Neuh{\"a}user}, {Haberl},
  {Hohle}, \& {Schwope}}]{2009A&A...497L...9H}
{Hambaryan}, V., {Neuh{\"a}user}, R., {Haberl}, F., {Hohle}, M.~M., \&
  {Schwope}, A.~D. 2009, \aap, 497, L9

\bibitem[{{Hambaryan} {et~al.}(2017){Hambaryan}, {Suleimanov}, {Haberl},
  {Schwope}, {Neuh{\"a}user}, {Hohle}, \& {Werner}}]{2017A&A...601A.108H}
{Hambaryan}, V., {Suleimanov}, V., {Haberl}, F., {et~al.} 2017, \aap, 601, A108

\bibitem[{{Hambaryan} {et~al.}(2011){Hambaryan}, {Suleimanov}, {Schwope},
  {Neuh{\"a}user}, {Werner}, \& {Potekhin}}]{2011A&A...534A..74H}
{Hambaryan}, V., {Suleimanov}, V., {Schwope}, A.~D., {et~al.} 2011, \aap, 534,
  A74

\bibitem[{{Heyl} \& {Kulkarni}(1998)}]{1998ApJ...506L..61H}
{Heyl}, J.~S. \& {Kulkarni}, S.~R. 1998, \apjl, 506, L61

\bibitem[{{Ho} {et~al.}(2007){Ho}, {Kaplan}, {Chang}, {van Adelsberg}, \&
  {Potekhin}}]{2007MNRAS.375..821H}
{Ho}, W.~C.~G., {Kaplan}, D.~L., {Chang}, P., {van Adelsberg}, M., \&
  {Potekhin}, A.~Y. 2007, \mnras, 375, 821

\bibitem[{{Ho} {et~al.}(2008){Ho}, {Potekhin}, \&
  {Chabrier}}]{2008ApJS..178..102H}
{Ho}, W.~C.~G., {Potekhin}, A.~Y., \& {Chabrier}, G. 2008, \apjs, 178, 102

\bibitem[{{Hohle} {et~al.}(2012){Hohle}, {Haberl}, {Vink}, {de Vries}, \&
  {Neuh{\"a}user}}]{2012MNRAS.419.1525H}
{Hohle}, M.~M., {Haberl}, F., {Vink}, J., {de Vries}, C.~P., \&
  {Neuh{\"a}user}, R. 2012, \mnras, 419, 1525

\bibitem[{{Jansen} {et~al.}(2001){Jansen}, {Lumb}, {Altieri}, {Clavel}, {Ehle},
  {Erd}, {Gabriel}, {Guainazzi}, {Gondoin}, {Much}, {Munoz}, {Santos},
  {Schartel}, {Texier}, \& {Vacanti}}]{2001A&A...365L...1J}
{Jansen}, F., {Lumb}, D., {Altieri}, B., {et~al.} 2001, \aap, 365, L1

\bibitem[{{Jethwa} {et~al.}(2015){Jethwa}, {Saxton}, {Guainazzi},
  {Rodriguez-Pascual}, \& {Stuhlinger}}]{2015A&A...581A.104J}
{Jethwa}, P., {Saxton}, R., {Guainazzi}, M., {Rodriguez-Pascual}, P., \&
  {Stuhlinger}, M. 2015, \aap, 581, A104

\bibitem[{{Kalberla} {et~al.}(2005){Kalberla}, {Burton}, {Hartmann}, {Arnal},
  {Bajaja}, {Morras}, \& {P{\"o}ppel}}]{2005A&A...440..775K}
{Kalberla}, P.~M.~W., {Burton}, W.~B., {Hartmann}, D., {et~al.} 2005, \aap,
  440, 775

\bibitem[{{Kaplan}(2008)}]{2008AIPC..983..331K}
{Kaplan}, D.~L. 2008, in American Institute of Physics Conference Series, Vol.
  983, 40 Years of Pulsars: Millisecond Pulsars, Magnetars and More, ed.
  C.~{Bassa}, Z.~{Wang}, A.~{Cumming}, \& V.~M. {Kaspi}, 331--339

\bibitem[{{Kaplan} {et~al.}(2011){Kaplan}, {Kamble}, {van Kerkwijk}, \&
  {Ho}}]{2011ApJ...736..117K}
{Kaplan}, D.~L., {Kamble}, A., {van Kerkwijk}, M.~H., \& {Ho}, W.~C.~G. 2011,
  \apj, 736, 117

\bibitem[{{Kaplan} \& {van Kerkwijk}(2009)}]{2009ApJ...705..798K}
{Kaplan}, D.~L. \& {van Kerkwijk}, M.~H. 2009, \apj, 705, 798

\bibitem[{{Kaspi} \& {Beloborodov}(2017)}]{2017ARA&A..55..261K}
{Kaspi}, V.~M. \& {Beloborodov}, A.~M. 2017, \araa, 55, 261

\bibitem[{{Lasker} {et~al.}(2008){Lasker}, {Lattanzi}, {McLean}, {Bucciarelli},
  {Drimmel}, {Garcia}, {Greene}, {Guglielmetti}, {Hanley}, {Hawkins},
  {Laidler}, {Loomis}, {Meakes}, {Mignani}, {Morbidelli}, {Morrison},
  {Pannunzio}, {Rosenberg}, {Sarasso}, {Smart}, {Spagna}, {Sturch},
  {Volpicelli}, {White}, {Wolfe}, \& {Zacchei}}]{2008AJ....136..735L}
{Lasker}, B.~M., {Lattanzi}, M.~G., {McLean}, B.~J., {et~al.} 2008, \aj, 136,
  735

\bibitem[{{Manchester} {et~al.}(2005){Manchester}, {Hobbs}, {Teoh}, \&
  {Hobbs}}]{2005AJ....129.1993M}
{Manchester}, R.~N., {Hobbs}, G.~B., {Teoh}, A., \& {Hobbs}, M. 2005, \aj, 129,
  1993

\bibitem[{{Motch} {et~al.}(1999){Motch}, {Haberl}, {Zickgraf}, {Hasinger}, \&
  {Schwope}}]{1999A&A...351..177M}
{Motch}, C., {Haberl}, F., {Zickgraf}, F.-J., {Hasinger}, G., \& {Schwope},
  A.~D. 1999, \aap, 351, 177

\bibitem[{{Motch} {et~al.}(2005){Motch}, {Sekiguchi}, {Haberl}, {Zavlin},
  {Schwope}, \& {Pakull}}]{2005A&A...429..257M}
{Motch}, C., {Sekiguchi}, K., {Haberl}, F., {et~al.} 2005, \aap, 429, 257

\bibitem[{{Motch} {et~al.}(2003){Motch}, {Zavlin}, \&
  {Haberl}}]{2003A&A...408..323M}
{Motch}, C., {Zavlin}, V.~E., \& {Haberl}, F. 2003, \aap, 408, 323

\bibitem[{{Ostriker} \& {Gunn}(1969)}]{1969ApJ...157.1395O}
{Ostriker}, J.~P. \& {Gunn}, J.~E. 1969, \apj, 157, 1395

\bibitem[{{{\"O}zel} \& {Freire}(2016)}]{2016ARA&A..54..401O}
{{\"O}zel}, F. \& {Freire}, P. 2016, \araa, 54, 401

\bibitem[{{Page}(1995)}]{1995ApJ...442..273P}
{Page}, D. 1995, \apj, 442, 273

\bibitem[{{Pavlov} {et~al.}(1995){Pavlov}, {Shibanov}, {Zavlin}, \&
  {Meyer}}]{1995ASIC..450...71P}
{Pavlov}, G.~G., {Shibanov}, Y.~A., {Zavlin}, V.~E., \& {Meyer}, R.~D. 1995, in
  NATO Advanced Science Institutes (ASI) Series C, Vol. 450, NATO Advanced
  Science Institutes (ASI) Series C, ed. M.~A. {Alpar}, U.~{Kiziloglu}, \&
  J.~{van Paradijs}, 71

\bibitem[{{Pavlov} {et~al.}(1999){Pavlov}, {Zavlin}, \&
  {Tr{\"u}mper}}]{1999ApJ...511L..45P}
{Pavlov}, G.~G., {Zavlin}, V.~E., \& {Tr{\"u}mper}, J. 1999, \apjl, 511, L45

\bibitem[{{P{\'e}rez-Azor{\'{\i}}n} {et~al.}(2006){P{\'e}rez-Azor{\'{\i}}n},
  {Miralles}, \& {Pons}}]{2006A&A...451.1009P}
{P{\'e}rez-Azor{\'{\i}}n}, J.~F., {Miralles}, J.~A., \& {Pons}, J.~A. 2006,
  \aap, 451, 1009

\bibitem[{{Perna} \& {Pons}(2011)}]{2011ApJ...727L..51P}
{Perna}, R. \& {Pons}, J.~A. 2011, \apjl, 727, L51

\bibitem[{{Perna} {et~al.}(2013){Perna}, {Vigan{\`o}}, {Pons}, \&
  {Rea}}]{2013MNRAS.434.2362P}
{Perna}, R., {Vigan{\`o}}, D., {Pons}, J.~A., \& {Rea}, N. 2013, \mnras, 434,
  2362

\bibitem[{{Pires} {et~al.}(2014){Pires}, {Haberl}, {Zavlin}, {Motch}, {Zane},
  \& {Hohle}}]{2014A&A...563A..50P}
{Pires}, A.~M., {Haberl}, F., {Zavlin}, V.~E., {et~al.} 2014, \aap, 563, A50

\bibitem[{{Pons} {et~al.}(2009){Pons}, {Miralles}, \&
  {Geppert}}]{2009A&A...496..207P}
{Pons}, J.~A., {Miralles}, J.~A., \& {Geppert}, U. 2009, \aap, 496, 207

\bibitem[{{Popov} {et~al.}(2010){Popov}, {Pons}, {Miralles}, {Boldin}, \&
  {Posselt}}]{2010MNRAS.401.2675P}
{Popov}, S.~B., {Pons}, J.~A., {Miralles}, J.~A., {Boldin}, P.~A., \&
  {Posselt}, B. 2010, \mnras, 401, 2675

\bibitem[{{Posselt} {et~al.}(2018){Posselt}, {Pavlov}, {Ertan}, {{\c
  C}al{\i}{\c s}kan}, {Luhman}, \& {Williams}}]{2018ApJ...865....1P}
{Posselt}, B., {Pavlov}, G.~G., {Ertan}, {\"U}., {et~al.} 2018, \apj, 865, 1

\bibitem[{{Posselt} {et~al.}(2014){Posselt}, {Pavlov}, {Popov}, \&
  {Wachter}}]{2014ApJS..215....3P}
{Posselt}, B., {Pavlov}, G.~G., {Popov}, S., \& {Wachter}, S. 2014, The
  Astrophysical Journal Supplement Series, 215, 3

\bibitem[{{Posselt} {et~al.}(2007){Posselt}, {Popov}, {Haberl}, {Tr{\"u}mper},
  {Turolla}, \& {Neuh{\"a}user}}]{2007Ap&SS.308..171P}
{Posselt}, B., {Popov}, S.~B., {Haberl}, F., {et~al.} 2007, \apss, 308, 171

\bibitem[{{Potekhin} {et~al.}(2015){Potekhin}, {De Luca}, \&
  {Pons}}]{2015SSRv..191..171P}
{Potekhin}, A.~Y., {De Luca}, A., \& {Pons}, J.~A. 2015, \ssr, 191, 171

\bibitem[{{Read} {et~al.}(2014){Read}, {Guainazzi}, \&
  {Sembay}}]{2014A&A...564A..75R}
{Read}, A.~M., {Guainazzi}, M., \& {Sembay}, S. 2014, \aap, 564, A75

\bibitem[{{Rodr{\'{\i}}guez Castillo} {et~al.}(2016){Rodr{\'{\i}}guez
  Castillo}, {Israel}, {Tiengo}, {Salvetti}, {Turolla}, {Zane}, {Rea},
  {Esposito}, {Mereghetti}, {Perna}, {Stella}, {Pons}, {Campana}, {G{\"o}tz},
  \& {Motta}}]{2016MNRAS.456.4145R}
{Rodr{\'{\i}}guez Castillo}, G.~A., {Israel}, G.~L., {Tiengo}, A., {et~al.}
  2016, \mnras, 456, 4145

\bibitem[{{Schwope} {et~al.}(2005){Schwope}, {Hambaryan}, {Haberl}, \&
  {Motch}}]{2005A&A...441..597S}
{Schwope}, A.~D., {Hambaryan}, V., {Haberl}, F., \& {Motch}, C. 2005, \aap,
  441, 597

\bibitem[{{Skrutskie} {et~al.}(2006){Skrutskie}, {Cutri}, {Stiening},
  {Weinberg}, {Schneider}, {Carpenter}, {Beichman}, {Capps}, {Chester},
  {Elias}, {Huchra}, {Liebert}, {Lonsdale}, {Monet}, {Price}, {Seitzer},
  {Jarrett}, {Kirkpatrick}, {Gizis}, {Howard}, {Evans}, {Fowler}, {Fullmer},
  {Hurt}, {Light}, {Kopan}, {Marsh}, {McCallon}, {Tam}, {Van Dyk}, \&
  {Wheelock}}]{2006AJ....131.1163S}
{Skrutskie}, M.~F., {Cutri}, R.~M., {Stiening}, R., {et~al.} 2006, \aj, 131,
  1163

\bibitem[{{Str{\"u}der} {et~al.}(2001){Str{\"u}der}, {Briel}, {Dennerl},
  {Hartmann}, {Kendziorra}, {Meidinger}, {Pfeffermann}, {Reppin}, {Aschenbach},
  {Bornemann}, {Br{\"a}uninger}, {Burkert}, {Elender}, {Freyberg}, {Haberl},
  {Hartner}, {Heuschmann}, {Hippmann}, {Kastelic}, {Kemmer}, {Kettenring},
  {Kink}, {Krause}, {M{\"u}ller}, {Oppitz}, {Pietsch}, {Popp}, {Predehl},
  {Read}, {Stephan}, {St{\"o}tter}, {Tr{\"u}mper}, {Holl}, {Kemmer}, {Soltau},
  {St{\"o}tter}, {Weber}, {Weichert}, {von Zanthier}, {Carathanassis}, {Lutz},
  {Richter}, {Solc}, {B{\"o}ttcher}, {Kuster}, {Staubert}, {Abbey}, {Holland},
  {Turner}, {Balasini}, {Bignami}, {La Palombara}, {Villa}, {Buttler},
  {Gianini}, {Lain{\'e}}, {Lumb}, \& {Dhez}}]{2001A&A...365L..18S}
{Str{\"u}der}, L., {Briel}, U., {Dennerl}, K., {et~al.} 2001, \aap, 365, L18

\bibitem[{{Suleimanov} {et~al.}(2010){Suleimanov}, {Hambaryan}, {Potekhin},
  {van Adelsberg}, {Neuh{\"a}user}, \& {Werner}}]{2010A&A...522A.111S}
{Suleimanov}, V., {Hambaryan}, V., {Potekhin}, A.~Y., {et~al.} 2010, \aap, 522,
  A111

\bibitem[{{Suleimanov} {et~al.}(2017){Suleimanov}, {Klochkov}, {Poutanen}, \&
  {Werner}}]{2017A&A...600A..43S}
{Suleimanov}, V.~F., {Klochkov}, D., {Poutanen}, J., \& {Werner}, K. 2017,
  \aap, 600, A43

\bibitem[{{Tetzlaff} {et~al.}(2010){Tetzlaff}, {Neuh{\"a}user}, {Hohle}, \&
  {Maciejewski}}]{2010MNRAS.402.2369T}
{Tetzlaff}, N., {Neuh{\"a}user}, R., {Hohle}, M.~M., \& {Maciejewski}, G. 2010,
  \mnras, 402, 2369

\bibitem[{{Tetzlaff} {et~al.}(2012){Tetzlaff}, {Schmidt}, {Hohle}, \&
  {Neuh{\"a}user}}]{2012PASA...29...98T}
{Tetzlaff}, N., {Schmidt}, J.~G., {Hohle}, M.~M., \& {Neuh{\"a}user}, R. 2012,
  \pasa, 29, 98

\bibitem[{{Thompson} \& {Duncan}(1995)}]{1995MNRAS.275..255T}
{Thompson}, C. \& {Duncan}, R.~C. 1995, \mnras, 275, 255

\bibitem[{{Thompson} \& {Duncan}(1996)}]{1996ApJ...473..322T}
{Thompson}, C. \& {Duncan}, R.~C. 1996, \apj, 473, 322

\bibitem[{{Tiengo} {et~al.}(2013){Tiengo}, {Esposito}, {Mereghetti}, {Turolla},
  {Nobili}, {Gastaldello}, {G{\"o}tz}, {Israel}, {Rea}, {Stella}, {Zane}, \&
  {Bignami}}]{2013Natur.500..312T}
{Tiengo}, A., {Esposito}, P., {Mereghetti}, S., {et~al.} 2013, \nat, 500, 312

\bibitem[{{Turner} {et~al.}(2001){Turner}, {Abbey}, {Arnaud}, {Balasini},
  {Barbera}, {Belsole}, {Bennie}, {Bernard}, {Bignami}, {Boer}, {Briel},
  {Butler}, {Cara}, {Chabaud}, {Cole}, {Collura}, {Conte}, {Cros}, {Denby},
  {Dhez}, {Di Coco}, {Dowson}, {Ferrando}, {Ghizzardi}, {Gianotti}, {Goodall},
  {Gretton}, {Griffiths}, {Hainaut}, {Hochedez}, {Holland}, {Jourdain},
  {Kendziorra}, {Lagostina}, {Laine}, {La Palombara}, {Lortholary}, {Lumb},
  {Marty}, {Molendi}, {Pigot}, {Poindron}, {Pounds}, {Reeves}, {Reppin},
  {Rothenflug}, {Salvetat}, {Sauvageot}, {Schmitt}, {Sembay}, {Short},
  {Spragg}, {Stephen}, {Str{\"u}der}, {Tiengo}, {Trifoglio}, {Tr{\"u}mper},
  {Vercellone}, {Vigroux}, {Villa}, {Ward}, {Whitehead}, \&
  {Zonca}}]{2001A&A...365L..27T}
{Turner}, M.~J.~L., {Abbey}, A., {Arnaud}, M., {et~al.} 2001, \aap, 365, L27

\bibitem[{{Turolla}(2009)}]{2009ASSL..357..141T}
{Turolla}, R. 2009, in Astrophysics and Space Science Library, Vol. 357,
  Astrophysics and Space Science Library, ed. W.~{Becker}, 141

\bibitem[{{Turolla} {et~al.}(2004){Turolla}, {Zane}, \&
  {Drake}}]{2004ApJ...603..265T}
{Turolla}, R., {Zane}, S., \& {Drake}, J.~J. 2004, \apj, 603, 265

\bibitem[{{Turolla} {et~al.}(2011){Turolla}, {Zane}, {Pons}, {Esposito}, \&
  {Rea}}]{2011ApJ...740..105T}
{Turolla}, R., {Zane}, S., {Pons}, J.~A., {Esposito}, P., \& {Rea}, N. 2011,
  \apj, 740, 105

\bibitem[{{Turolla} {et~al.}(2015){Turolla}, {Zane}, \&
  {Watts}}]{2015RPPh...78k6901T}
{Turolla}, R., {Zane}, S., \& {Watts}, A.~L. 2015, Reports on Progress in
  Physics, 78, 116901

\bibitem[{{van Kerkwijk} \& {Kaplan}(2007)}]{2007Ap&SS.308..191V}
{van Kerkwijk}, M.~H. \& {Kaplan}, D.~L. 2007, \apss, 308, 191

\bibitem[{{van Kerkwijk} {et~al.}(2004){van Kerkwijk}, {Kaplan}, {Durant},
  {Kulkarni}, \& {Paerels}}]{2004ApJ...608..432V}
{van Kerkwijk}, M.~H., {Kaplan}, D.~L., {Durant}, M., {Kulkarni}, S.~R., \&
  {Paerels}, F. 2004, \apj, 608, 432

\bibitem[{{Vigan{\`o}} {et~al.}(2014){Vigan{\`o}}, {Perna}, {Rea}, \&
  {Pons}}]{2014MNRAS.443...31V}
{Vigan{\`o}}, D., {Perna}, R., {Rea}, N., \& {Pons}, J.~A. 2014, \mnras, 443,
  31

\bibitem[{{Vigan{\`o}} {et~al.}(2013){Vigan{\`o}}, {Rea}, {Pons}, {Perna},
  {Aguilera}, \& {Miralles}}]{2013MNRAS.434..123V}
{Vigan{\`o}}, D., {Rea}, N., {Pons}, J.~A., {et~al.} 2013, \mnras, 434, 123

\bibitem[{{Voges} {et~al.}(1999){Voges}, {Aschenbach}, {Boller},
  {Br{\"a}uninger}, {Briel}, {Burkert}, {Dennerl}, {Englhauser}, {Gruber},
  {Haberl}, {Hartner}, {Hasinger}, {K{\"u}rster}, {Pfeffermann}, {Pietsch},
  {Predehl}, {Rosso}, {Schmitt}, {Tr{\"u}mper}, \&
  {Zimmermann}}]{1999A&A...349..389V}
{Voges}, W., {Aschenbach}, B., {Boller}, T., {et~al.} 1999, \aap, 349, 389

\bibitem[{{Walter} {et~al.}(2010){Walter}, {Eisenbei{\ss}}, {Lattimer}, {Kim},
  {Hambaryan}, \& {Neuh{\"a}user}}]{2010ApJ...724..669W}
{Walter}, F.~M., {Eisenbei{\ss}}, T., {Lattimer}, J.~M., {et~al.} 2010, \apj,
  724, 669

\bibitem[{{Wang} {et~al.}(2017){Wang}, {Lu}, {Tong}, {Ge}, {Li}, {Men}, \&
  {Xu}}]{2017ApJ...837...81W}
{Wang}, W., {Lu}, J., {Tong}, H., {et~al.} 2017, \apj, 837, 81

\bibitem[{{Willingale} {et~al.}(2013){Willingale}, {Starling}, {Beardmore},
  {Tanvir}, \& {O'Brien}}]{2013MNRAS.431..394W}
{Willingale}, R., {Starling}, R.~L.~C., {Beardmore}, A.~P., {Tanvir}, N.~R., \&
  {O'Brien}, P.~T. 2013, \mnras, 431, 394

\bibitem[{{Wilms} {et~al.}(2000){Wilms}, {Allen}, \&
  {McCray}}]{2000ApJ...542..914W}
{Wilms}, J., {Allen}, A., \& {McCray}, R. 2000, \apj, 542, 914

\bibitem[{{Zane} {et~al.}(2006){Zane}, {de Luca}, {Mignani}, \&
  {Turolla}}]{2006A&A...457..619Z}
{Zane}, S., {de Luca}, A., {Mignani}, R.~P., \& {Turolla}, R. 2006, \aap, 457,
  619

\bibitem[{{Zane} \& {Turolla}(2006)}]{2006MNRAS.366..727Z}
{Zane}, S. \& {Turolla}, R. 2006, \mnras, 366, 727

\bibitem[{{Zavlin} {et~al.}(1996){Zavlin}, {Pavlov}, \&
  {Shibanov}}]{1996A&A...315..141Z}
{Zavlin}, V.~E., {Pavlov}, G.~G., \& {Shibanov}, Y.~A. 1996, \aap, 315, 141

\end{thebibliography}
\end{document}